\documentstyle[aps,manuscript,psfig]{revtex}
\begin{document}
\draft
\renewcommand{\thefootnote}{\fnsymbol{footnote}}
\begin{title}
{\bf Current Dissipation in Thin Superconducting Wires: \\
Accurate Numerical Evaluation Using the String Method}
\end{title}
\author{Tiezheng Qian}
\address{Department of Mathematics,
Hong Kong University of Science and Technology,\\
Clear Water Bay, Kowloon, Hong Kong, China}
\author{Weiqing Ren}
\address{Department of Mathematics,Princeton University,
Princeton, New Jersey 08544, USA}
\author{Ping Sheng}
\address{Department of Physics and Institute of Nano Science and Technology,\\
Hong Kong University of Science and Technology,
Clear Water Bay, Kowloon, Hong Kong, China}
\maketitle
 
\begin{abstract}
Current dissipation in thin superconducting wires is numerically evaluated
by using the string method, within the framework of time-dependent 
Ginzburg-Landau equation with a Langevin noise term.
The most probable transition pathway between two neighboring
current-carrying metastable states, continuously linking 
the Langer-Ambegaokar saddle-point state to a state
in which the order parameter vanishes somewhere, is found numerically. 
We also give a numerically accurate algorithm to evaluate the prefactors 
for the rate of current-reducing transitions.
\end{abstract}
\narrowtext

\section{Introduction}\label{intro}
The picture of resistive (current-reducing) phase slips was first 
discussed by Little \cite{little}. Langer and Ambegaokar then used
a Ginzburg-Landau free-energy functional to analytically obtain
the lowest free-energy saddle point between 
two current-carrying metastable states \cite{LA}. The time scale of 
the resistive phase slips have been formulated by McCumber and Halperin 
\cite{MH}. The theory developed in Refs. \cite{LA} and \cite{MH}
is generally referred to as 
the Langer-Ambegaokar-McCumber-Halperin (LAMH) theory. Recently, 
new technique has been developed for fabricating superconducting
nanowires. Since resistive transition region broadens with decreasing 
cross-sectional area of the wire, nanowires therefore become 
ideal samples for a more precise test of the LAMH theory \cite{tinkham}.
For this purpose, a quantitative evaluation of the thermal-activation rate
of phase slip events is needed. In particular, this is the case since
the thermal rate serves as the background for distinguishing quantum fluctuations at low temperature, a topic of considerable basic scientific interest \cite{tinkham}

Recently, the string method \cite{ren-prb,ren-jap,ERE,Ren} has been 
presented for the numerical evaluation of thermally activated rare events. 
This method first locates the most probable transition pathway 
connecting two metastable states in configuration space.
This is done by evolving strings, i.e., smooth curves with intrinsic
parametrization, into the minimal energy path. The transition rates can 
then be computed using an umbrella sampling technique which simulates
the fluctuations around the most probable path.
In this paper we show that the string method can be employed as
an efficient numerical tool for the study of thermally activated 
phase slips in thin superconducting wires below $T_c$.

The system is modeled by a one-dimensional (1D) time-dependent 
Ginzburg-Landau equation (TDGLE) with a Langevin noise term.
Applying the string method to this particular system, 
we obtain the most probable transition pathway between two neighboring
current-carrying metastable states. This pathway continuously
connects the Langer-Ambegaokar saddle-point state \cite{LA} to a state
in which the order parameter vanishes somewhere to allow
a phase slip of $2\pi$, as first proposed by Little \cite{little}.
We also give a numerically accurate algorithm to evaluate the prefactors 
for the rate of resistive phase slips.

\section{String Method}\label{string}
To outline the string method \cite{ren-prb}, 
consider a system governed by the overdamped Langevin equation
\begin{equation}\label{langevin-eq}
\gamma\dot{q_i}=-\nabla_i V({\bf q})+\zeta_i(t),
\end{equation}
where $\gamma$ is the frictional coefficient, $\bf q$ denotes 
the generalized coordinates $\{q_i\}$, 
$\dot{q_i}=\partial q_i/\partial t$, $\nabla_i=\partial/\partial q_i$,
and $\zeta_i(t)$ is a white noise satisfying
$\langle\zeta_i(t)\zeta_j(t')\rangle=
2\gamma k_BT\delta_{ij}\delta(t-t')$, with $k_B$ denoting 
the Boltzmann constant and $T$ the temperature.
Metastable and stable states are
located in configuration space as the minima of the potential 
$V({\bf q})$. Assume ${\bf q}_A$ and ${\bf q}_B$ are the two minima 
of $V$. In terms of the topography of $V({\bf q})$, 
the most probable fluctuation which can carry the system from 
${\bf q}_A$ to ${\bf q}_B$ (or ${\bf q}_B$ to ${\bf q}_A$)
corresponds to the lowest intervening saddle point between 
these two minima. The minimal energy path (MEP) is defined as 
a smooth curve ${\bf q}^*(s)$ connecting ${\bf q}_A$ and ${\bf q}_B$ 
with intrinsic parametrization such as arc length $s$, 
which satisfies
\begin{equation}\label{mep-def}
\left(\nabla V\right)^\perp({\bf q}^*)=0,
\end{equation}
where $\left(\nabla V\right)^\perp$ is the component of $\nabla V$ 
normal to the path ${\bf q}^*(s)$. This MEP is the most probable 
pathway for thermally activated transitions between ${\bf q}_A$ and 
${\bf q}_B$. To numerically locate the MEP in configuration space, 
a string ${\bf q}(s)$ (a smooth curve with intrinsic parametrization 
by $s$) connecting ${\bf q}_A$ and ${\bf q}_B$ is evolved according to
\begin{equation}\label{string-evolution}
\dot{\bf q}=-\left(\nabla V\right)^\perp({\bf q}).
\end{equation}
A re-parametrization is applied once in a while to enforce accurate
parametrization by arc length. The stationary solution
of Eq. (\ref{string-evolution}) satisfies Eq. (\ref{mep-def})
which defines the MEP.

Once the MEP is determined, the lowest saddle point is known and
the transition rate can be computed by evaluating the fluctuations 
around the MEP \cite{ren-prb}. 
Following Kramers' approach and its generalizations 
\cite{kramers,landauer-swanson,langer},
the transition rate is given by
\begin{equation}\label{thermal-rate}
\Gamma_T({A\rightarrow B})=\displaystyle\frac{|\lambda_s|}{2\pi\gamma}
\left[\displaystyle\frac{\det H({\bf q}_A)}{|\det H({\bf q}_s)|}
\right]^{1/2}
\exp\left\{-\displaystyle\frac{1}{k_BT}
[V({\bf q}_s)-V({\bf q}_A)]\right\},
\end{equation}
where ${\bf q}_s$ is the saddle point found at the MEP, $H({\bf q})$
denotes the Hessian of $V({\bf q})$, and $\lambda_s$ is the negative
eigenvalue of $H({\bf q}_s)$. (By definition, $H({\bf q}_s)$ has one 
and only one negative eigenvalue.) The determinant ratio in 
Eq. (\ref{thermal-rate}) are numerically obtained by 
linear interpolation as follows \cite{ERE}.

Let $F$ and $G$ be two $N\times N$ positive definite matrices and 
$\bf q$ a column vector in $R^N$ space. A harmonic potential
parametrized by $\alpha$ ($0\le \alpha\le 1$) is constructed as
\begin{equation}\label{potential-U}
U^\alpha({\bf q})=\displaystyle\frac{1}{2}
{\bf q}^T[(1-\alpha)F+\alpha G]{\bf q},
\end{equation}
with the corresponding partition function given by
\begin{equation}\label{partition-Z}
Z(\alpha)=\int d{\bf q}\exp\left[-\displaystyle\frac{1}{\epsilon}
U^\alpha({\bf q})\right]=(2\pi\epsilon)^{N/2}
\left(\det[(1-\alpha)F+\alpha G]\right)^{-1/2}.
\end{equation}
From the expectation value
\begin{equation}\label{expecta}
\begin{array}{ll}
\displaystyle\frac{d}{d\alpha}\ln Z(\alpha) & 
=\displaystyle\frac{1}{Z(\alpha)}\int d{\bf q}
\left[\displaystyle\frac{1}{2\epsilon}{\bf q}^T(F-G){\bf q}\right]
\exp\left[-\displaystyle\frac{1}{\epsilon}U^\alpha({\bf q})\right]\\
& =\left\langle
\left[\displaystyle\frac{1}{2\epsilon}{\bf q}^T(F-G){\bf q}\right]
\right\rangle_\alpha :=Q(\alpha),
\end{array}
\end{equation}
we have
\begin{equation}\label{partion-ratio}
\displaystyle\frac{Z(1)}{Z(0)}=\exp\left\{\int_0^1 Q(\alpha) d\alpha\right\}.
\end{equation}
It follows from (\ref{partition-Z}) that
\begin{equation}\label{determinant-ratio}
\displaystyle\frac{\det F}{\det G}=\exp\left\{2\int_0^1 
Q(\alpha) d\alpha \right\}.
\end{equation}
The expectation value $Q(\alpha)$ can be 
numerically evaluated in the canonical ensemble governed by potential 
$U^\alpha({\bf q})$. In practice, the ensemble is generated by solving
\begin{equation}\label{eqn-sde}
\dot q=-\nabla U^\alpha(q) + \zeta,
\end{equation}
where $\zeta(t)$ is a white noise satisfying 
$\langle\zeta_i(t)\zeta_j(t')\rangle = 2\epsilon \delta_{ij}\delta(t-t')$.

To apply the above technique to the present problem, it is noted that
the Hessian at the saddle point, $H({\bf q}_s)$, 
has a negative eigenvalue $\lambda_s$. Given this $\lambda_s$ and 
the corresponding normalized eigenvector ${\bf u}_s$, the indefinite 
$H({\bf q}_s)$ has to be modified to give a positive definite 
$\tilde{H}({\bf q}_s)$:
\begin{equation}\label{positive-definite}
\tilde{H}({\bf q}_s)=H({\bf q}_s)+(\nu-\lambda_s)
{\bf u}_s{\bf u}_s^T,
\end{equation}
where $\nu$ is a positive parameter. It follows that
$\det \tilde{H}({\bf q}_s)$ and $\det H({\bf q}_s)$ are related by
$$\det \tilde{H}({\bf q}_s)=\displaystyle\frac{\nu}{\lambda_s}
\det H({\bf q}_s),$$
if we remember that the determinant is the product of the eigenvalues.
From the MEP ${\bf q}^*(s)$ parametrized by the arc length $s$,
the eigenvector ${\bf u}_s$ can be obtained by evaluating 
$d{\bf q}^*(s)/ds$ at the saddle point, followed by a normalization, 
and $\lambda_s$ is then computed from 
$\lambda_s=[{\bf u}_s]^TH({\bf q}_s){\bf u}_s$. The ratio 
${\det H({\bf q}_A)}/{\det \tilde{H}({\bf q}_s)}$ can be readily 
computed according to Eq. (\ref{determinant-ratio}) because 
$H({\bf q}_A)$ and $\tilde{H}({\bf q}_s)$ are both positive definite.
The determinant ratio in the rate expression (\ref{thermal-rate}) 
is then obtained from 
$$\displaystyle\frac{\det H({\bf q}_A)}{\det H({\bf q}_s)}=
\displaystyle\frac{\nu}{\lambda_s}
\displaystyle\frac{\det H({\bf q}_A)}{\det \tilde{H}({\bf q}_s)}.$$

\section{Phase-Slip Fluctuations in One-Dimensional Superconductor}
\label{physics}
\subsection{One-Dimensional Superconductor}\label{superconductor}
For a superconducting wire below $T_c$, if the transverse dimension
$d\ll$ the coherence length $\xi(T)$, then the variations of 
the order parameter $\psi$ over the cross section of the wire are
energetically prohibited. The wire sample therefore becomes a 1D
superconductor, with $\psi$ being a function of a single coordinate 
$x$ along the wire. The Ginzburg-Landau free-energy functional
is of the form
\begin{equation}\label{GL-free-energy}
F[\psi(x)]=\sigma\int dx\left[
\displaystyle\frac{K}{2}|\nabla\psi(x)|^2-
\displaystyle\frac{\alpha_0(T_c-T)}{2}|\psi(x)|^2+
\displaystyle\frac{\beta}{4}|\psi(x)|^4\right],
\end{equation}
where $\sigma$ is the cross-sectional area of the wire,  
$K=\hbar^2/m^*$ with $m^*$ the effective mass of the Cooper pair,
and $\alpha_0$ and $\beta$ are both phenomenological parameters.
The time evolution of $\psi$ is governed by the time-dependent
Ginzburg-Landau equation (TDGLE)
\begin{equation}\label{TDGLE}
\gamma\displaystyle\frac{\partial}{\partial t}{\psi}
=-\displaystyle\frac{1}{\sigma}\displaystyle\frac
{\delta F[\psi]}{\delta \psi}+\zeta
=K\nabla^2\psi+\alpha_0(T_c-T)\psi
-\beta|\psi|^2\psi+\zeta,
\end{equation}
where $\gamma$ is a viscosity coefficient, and $\zeta(x,t)$ is 
a Langevin white noise, with autocorrelation functions 
$$\langle\zeta(x,t)\zeta(x',t')\rangle=0;\;\;\;
\langle\zeta(x,t)\zeta^*(x',t')\rangle=
4\sigma^{-1}\gamma k_BT\delta(x-x')\delta(t-t').$$ 
This noise generates a random motion of $\psi$ and stabilizes 
the equilibrium distribution, which is proportional to 
$\exp\left\{-F[\psi(x)]/k_BT\right\}$. 

For the convenience of presentation and computation, we use
the dimensionless form 
\begin{equation}\label{GL-free-energy1}
\bar{F}[\bar{\psi}(\bar{x})]=\int d\bar{x}\left[
\displaystyle\frac{1}{2}|\bar{\nabla}\bar{\psi}(\bar{x})|^2-
\displaystyle\frac{1}{2}|\bar{\psi}(\bar{x})|^2+
\displaystyle\frac{1}{4}|\bar{\psi}(\bar{x})|^4\right],
\end{equation}
for the free-energy functional. Here the over bar
denotes the dimensionless quantities, obtained with 
$F$ scaled by $\sigma\xi\alpha_0^2(T_c-T)^2/\beta$,
$\psi$ by $\sqrt{\alpha_0(T_c-T)/\beta}$, and $x$ by 
the correlation length $\xi(T)=\sqrt{K/\alpha_0(T_c-T)}$.
Correspondingly, the dimensionless TDGLE is of the form
\begin{equation}\label{TDGLE1}
\displaystyle\frac{\partial}{\partial\bar{t}}\bar{\psi}=
-\displaystyle\frac
{\delta\bar{F}[\bar{\psi}]}{\delta\bar{\psi}}+\bar{\zeta}
=\bar{\nabla}^2\bar{\psi}+\bar{\psi}-|\bar{\psi}|^2\bar{\psi}
+\bar{\zeta},
\end{equation}
in which the time is scaled by $\tau(T)=\gamma/\alpha_0(T_c-T)$, 
and the dimensionless noise $\bar{\zeta}$ satisfies 
the autocorrelation functions
$$\langle\bar{\zeta}(\bar{x},\bar{t})
\bar{\zeta}(\bar{x}',\bar{t}')\rangle=0;\;\;\;
\langle\bar{\zeta}(\bar{x},\bar{t})
\bar{\zeta}^*(\bar{x}',\bar{t}')\rangle=
\displaystyle\frac{4k_BT}{\sigma\xi\alpha_0^2(T_c-T)^2/\beta}
\delta(\bar{x}-\bar{x}')\delta(\bar{t}-\bar{t}').$$ 
Throughout the remainder of this paper, all physical quantities 
are given in terms of the dimensionless quantities, using 
$\sigma\xi\alpha_0^2(T_c-T)^2/\beta=\sigma\xi{H_c^2(T)}/{2\pi}$ 
for the energy scale ($\alpha_0^2(T_c-T)^2/4\beta={H_c^2(T)}/{8\pi}$
is the condensation energy density, 
where $H_c(T)$ is the bulk critical field),
$\sqrt{\alpha_0(T_c-T)/\beta}$ for the $\psi$ scale,
$\xi$ for the length scale, and $\tau$ for the time scale.
Note that all the temperature effects are absorbed into 
these scales. The over bar will be dropped in the remainder of the paper.

\subsection{Current-Carrying Metastable States}\label{meta}
Consider a closed superconducting ring. 
Periodic boundary condition for $\psi(x)$ is imposed by
$\psi(-l/2)=\psi(l/2)$ where $l$ is the circumference of the ring.
Metastable current-carrying states $\psi_n$ are obtained 
from the stationary Ginzburg-Landau equation
\begin{equation}\label{stationaryGLE}
-{\nabla}^2{\psi}-{\psi}+|{\psi}|^2{\psi}=0,
\end{equation}
as local minima of $F$:
\begin{equation}\label{metastable}
\psi_n(x)=f_ne^{ik_nx}=\sqrt{1-k_n^2}e^{ik_nx},
\;\;\;k_n=2\pi n/l
\end{equation}
where $k_n$ is the wave vector, $f_n$ is the amplitude,
and $n$ is an integer. The dimensionless current density
in the $\psi_n$ state is given by $J_n=f_n^2k_n=(1-k_n^2)k_n$.
The metastability of $\psi_n$ requires $|k_n|<k_c=1/\sqrt{3}$.
In the presence of thermodynamic fluctuations, the lifetime
of these metastable states is finite. When the lifetime
is made sufficiently short, the decay of persistent current 
becomes observable.

\subsection{Current-Reducing Phase-Slip Fluctuations}\label{slip}
The decay of persistent current in a superconducting
ring may be explained by dividing the $\psi$-function space
into different subspaces, each labeled by an integer $n$, 
defined through $\phi(l/2)-\phi(-l/2)=2\pi n$ according to
the periodic boundary condition for $\psi=|\psi|e^{i\phi}$.
In each $\psi$-function subspace there is a current-carrying
metastable state $\psi_n$, defined as a local minimum of $F$. 

On the one hand, there are many low-energy configurations that 
are frequently accessed by the fluctuating system. Nevertheless, 
such low-energy fluctuations cause no change of the phase difference 
$\phi(l/2)-\phi(-l/2)$ across the whole ring, 
and therefore the global phase coherence persists.
On the other hand, there exist thermodynamic fluctuations that
lead to transitions between different $\psi$-function subspaces. 
These fluctuations involve large amplitude fluctuations of $\psi$. Heuristically, if $\psi$ vanishes somewhere, 
then $\phi(l/2)-\phi(-l/2)$ may change (slip) by $2\pi$
and hence the system moves from one subspace to another
(from $\psi_n$ to $\psi_{n+1}$ or $\psi_{n-1}$) \cite{little}.
Since larger persistent current means higher free energy,
transitions among different metastable states tend to reduce
the persistent current on average.
Large amplitude fluctuations usually cost free energies much higher 
than $k_BT$, therefore phase-slip events are rare. 
Only when the cross section of the ring is very small and 
the temperature is close enough to $T_c$, 
the decay of persistent currents due to infrequent phase-slip
fluctuations becomes observable.

\subsection{Free-Energy Saddle Point}\label{saddle-point}
Based on the Ginzburg-Landau free-energy functional $F$,
Langer and Ambegaokar have derived the lowest saddle point $\psi_s(x)$ 
between the two neighboring metastable states $\psi_n(x)$ and
$\psi_{n-1}(x)$ \cite{LA}.
This $\psi_s$ state corresponds to the most probable fluctuation
which can carry the system from $\psi_n$ to $\psi_{n-1}$
(or from $\psi_{n-1}$ to $\psi_n$). Analytical expressions for 
the saddle-point state $\psi_s$ and the dimensionless energy barriers
$\Delta F_-=F[\psi_s]-F[\psi_n]$ and 
$\Delta F_+=F[\psi_s]-F[\psi_{n-1}]$ are given in Appendix \ref{app_LA}.
In Sec. \ref{numerical}, we will show that using the string method,
$\psi_s$ and $\Delta F_{\pm}$ can be numerically obtained from the MEP
connecting $\psi_n(x)$ and $\psi_{n-1}(x)$.

\section{Application of the String Method: Fluctuation Time Scale}
\label{time-scale}
The time scale of the thermodynamic phase-slip fluctuations
is determined by the Langevin equation (\ref{TDGLE}).
The transition rates $\Gamma_{\pm}$ for the transitions 
$\psi_{n-1}\rightarrow\psi_n$ and $\psi_n\rightarrow\psi_{n-1}$
can be written as
\begin{equation}\label{MH-rate}
\Gamma_{\pm}=\Omega_{\pm}\exp\left[-\displaystyle\frac
{\sigma\xi H_c^2(T)}{2\pi k_BT}\Delta F_{\pm}\right],
\end{equation}
where $\Omega_{\pm}$ are the prefactors which fix the fluctuation 
time scale, and $\sigma\xi{H_c^2(T)}/{2\pi}$ is the energy unit.
Based upon Kramers' formulation and its generalizations 
\cite{kramers,landauer-swanson,langer}, McCumber and Halperin
have derived an analytical expression for the prefactors $\Omega_{\pm}$ 
\cite{MH}. To our knowledge, numerical evaluation of $\Omega_{\pm}$ 
has never been reported. Here we outline a numerical scheme for 
the evaluation of $\Omega_{\pm}$ so that a complete solution of
the LAMH theory may be obtained. Results based on this scheme will be
presented in Sec. \ref{numerical}.

We adopt a two-component vector representation for the complex $\psi$:
$${\mbox{\boldmath$\eta$}}(x)=\left[\eta_1(x)\;\;\eta_2(x)\right]^T=
\left[{\rm Re}\psi(x)\;\;{\rm Im}\psi(x)\right]^T.$$
In terms of ${\mbox{\boldmath$\eta$}}$, 
the dimensionless form of the free-energy functional
in Eq. (\ref{GL-free-energy1}) becomes
\begin{equation}\label{GL-free-energy2}
F[{\mbox{\boldmath$\eta$}}(x)]=\int dx\left[
\displaystyle\frac{1}{2}(\nabla{\mbox{\boldmath$\eta$}})^2-
\displaystyle\frac{1}{2}{\mbox{\boldmath$\eta$}}^2+
\displaystyle\frac{1}{4}({\mbox{\boldmath$\eta$}}^2)^2\right],
\end{equation}
and the dimensionless TDGLE becomes
\begin{equation}\label{TDGLE2}
\displaystyle\frac{\partial}{\partial t}{\mbox{\boldmath$\eta$}}
=-\displaystyle\frac{\delta}{\delta {\mbox{\boldmath$\eta$}}}
F[{\mbox{\boldmath$\eta$}}(x)]+{\mbox{\boldmath$\zeta$}}=
\nabla^2{\mbox{\boldmath$\eta$}}+{\mbox{\boldmath$\eta$}}
-{\mbox{\boldmath$\eta$}}^2{\mbox{\boldmath$\eta$}}
+{\mbox{\boldmath$\zeta$}},
\end{equation}
in which the noise ${\mbox{\boldmath$\zeta$}}(x)=
\left[\zeta_1(x)\;\;\zeta_2(x)\right]^T$ satisfies 
the autocorrelation functions
$$\langle\zeta_i(x,t)\zeta_j(x',t')\rangle=
\displaystyle\frac{2k_BT}{\sigma\xi\alpha_0^2(T_c-T)^2/\beta}
\delta_{ij}\delta(x-x')\delta(t-t').$$ 
The vector forms for the metastable state $\psi_n$
and the saddle-point state $\psi_s$ can be easily obtained. 
The Hessian of $F[{\mbox{\boldmath$\eta$}}(x)]$ is given by
\begin{equation}\label{hessian}
H[{\mbox{\boldmath$\eta$}}]=\left[\begin{array}{cc}
-\nabla^2-1+3\eta_1^2+\eta_2^2 & 2\eta_1\eta_2\\
2\eta_1\eta_2 & -\nabla^2-1+\eta_1^2+3\eta_2^2
\end{array}\right].
\end{equation}

The general expression (\ref{thermal-rate}) for 
the thermal-activation rate can be directly applied to 
the phase-slip fluctuations, with some elaboration for 
the symmetry properties of the system. According to
Eq. (\ref{thermal-rate}), the dimensionless form of
the prefactors $\Omega_{\pm}$ in Eq. (\ref{MH-rate}) 
can be formally written as
\begin{equation}\label{formal-omega}
\Omega_{+}=\displaystyle\frac{|\lambda_s^{(1)}|}{2\pi}
\left[\displaystyle\frac{\det H_{n-1}}{|\det H_s|}
\right]^{1/2},\;\;\;
\Omega_{-}=\displaystyle\frac{|\lambda_s^{(1)}|}{2\pi}
\left[\displaystyle\frac{\det H_n}{|\det H_s|}
\right]^{1/2},
\end{equation}
where $H_{n-1}$, $H_n$, and $H_s$ are the three Hessians
evaluated at $\psi_{n-1}$, $\psi_n$, and $\psi_s$ according to
$H[{\mbox{\boldmath$\eta$}}]$ in Eq. (\ref{hessian}), and 
$\lambda_s^{(1)}$ is the lowest (negative) eigenvalue of $H_s$.

The free energy $F$ is invariant under the gauge transformation
$\psi(x)\rightarrow\psi(x)e^{i\phi_0}$
and the translational transformation
$\psi(x)\rightarrow\psi(x-x_0)$. 
(The gauge invariance in terms of $\psi$ is equivalent to
the rotational invariance in terms of ${\mbox{\boldmath$\eta$}}$).
As a consequence, $H_n$ must have a zero eigenvalue 
$\lambda_n^{(1)}$ (the lowest one, 
other eigenvalues are all positive), 
with the corresponding eigenvector 
\begin{equation}\label{un1}
{\bf u}_n^{(1)}(x)=\alpha_n^{(1)}
\left[\begin{array}{cc}0 & -1 \\ 1 & 0 \end{array}\right]
{\mbox{\boldmath$\eta$}}_n(x),
\end{equation}
where ${\mbox{\boldmath$\eta$}}_n$ is the vector form for $\psi_n$
and $\alpha_n^{(1)}$ is the normalization factor.
Similarly, $H_s$ must have a zero eigenvalue $\lambda_s^{(2)}$, with
the corresponding eigenvector 
\begin{equation}\label{us2}
{\bf u}_s^{(2)}(x)=\alpha_s^{(2)}
\left[\begin{array}{cc}0 & -1 \\ 1 & 0 \end{array}\right]
{\mbox{\boldmath$\eta$}}_s(x),
\end{equation}
where ${\mbox{\boldmath$\eta$}}_s$ is the vector form for $\psi_s$
and $\alpha_s^{(2)}$ is the normalization factor.
The translational invariance of $F$ leads to another zero eigenvalue
$\lambda_s^{(3)}$ because the phase-slip center in the saddle-point state 
$\psi_s$ can be continuously shifted (see Eq. (\ref{barrier-state})).
%The eigenvector corresponding to $\lambda_s^{(3)}$ will be discussed later.

The presence of the zero eigenvalues $\lambda_n^{(1)}$,
$\lambda_s^{(2)}$, and $\lambda_s^{(3)}$ requires some extra efforts
in evaluating the prefactors in Eq. (\ref{formal-omega}).
While the gauge invariance comes from an unphysical degree
of freedom and hence $\lambda_n^{(1)}$ and $\lambda_s^{(2)}$
are simply discarded, the translational invariance, however,
points to the fact that phase-slip fluctuations in different
spatial regions are equally probable. The total rate for
a transition, say $\psi_{n-1}\rightarrow\psi_n$, should
be obtained by summing over all the phase-slip fluctuations
across the whole length of the system. This is achieved as follows.

The prefactors $\Omega_{\pm}$ in Eq. (\ref{formal-omega}) appear to
diverge because of the presence of $[\lambda_s^{(3)}]^{-1/2}$ 
in $(\det H_s)^{-1/2}$. This $[\lambda_s^{(3)}]^{-1/2}$ 
is contributed by the integral
\begin{equation}\label{zero-1}
\sqrt{\displaystyle\frac{\sigma\xi H_c^2(T)}{2\pi k_BT}}\int
\displaystyle\frac{dc_s^{(3)}}{\sqrt{2\pi}}\exp\left[
-\displaystyle\frac{\sigma\xi H_c^2(T)}{2\pi k_BT}\left(
\displaystyle\frac{1}{2}\lambda_s^{(3)}[c_s^{(3)}]^2\right)\right]=
\left[\lambda_s^{(3)}\right]^{-1/2},
\end{equation}
where $(1/2)\lambda_s^{(3)}[c_s^{(3)}]^2$ comes from 
a Taylor expansion at ${\mbox{\boldmath$\eta$}}_s$ for 
$F[{\mbox{\boldmath$\eta$}}]$, representing the second-order term 
contributed by the component of
${\mbox{\boldmath$\eta$}}-{\mbox{\boldmath$\eta$}}_s$
in the direction of ${\bf u}_s^{(3)}$.
Here ${\bf u}_s^{(3)}(x)$ is the normalized eigenvector
corresponding to $\lambda_s^{(3)}$ and $c_s^{(3)}$ is 
the coordinate in the direction of ${\bf u}_s^{(3)}(x)$:
$$H_s{\bf u}_s^{(3)}(x)=\lambda_s^{(3)}{\bf u}_s^{(3)}(x),\;\;\;
c_s^{(3)}=\int dx [{\bf u}_s^{(3)}(x)]^T\left[
{\mbox{\boldmath$\eta$}}(x)-{\mbox{\boldmath$\eta$}}_s(x)\right].$$
Given $\lambda_s^{(3)}=0$, equation (\ref{zero-1}) may be rewritten as 
\begin{equation}\label{zero-2}
\left[\lambda_s^{(3)}\right]^{-1/2}=\displaystyle\frac{1}{2\pi}
\sqrt{\displaystyle\frac{\sigma\xi H_c^2(T)}{k_BT}}
\int dc_s^{(3)}.
\end{equation}
It can be shown that $\int dc_s^{(3)}$ is an integral proportional to 
the system length $l$: $\int dc_s^{(3)}=\Lambda l$, and hence
\begin{equation}\label{zero-3}
\left[\lambda_s^{(3)}\right]^{-1/2}=\displaystyle\frac{1}{2\pi}
\sqrt{\displaystyle\frac{\sigma\xi H_c^2(T)}{k_BT}}\Lambda l.
\end{equation}
Thus the prefactors $\Omega_\pm$ in Eq. (\ref{formal-omega}) are 
proportional to the system length, as required physically 
by the translational symmetry of the system.
In Ref. \cite{MH} an analytical expression has been derived for $\Lambda$ 
(see Appendix. \ref{app_MH}). Below we outline a method for evaluating
$\Lambda$ numerically.

The eigenspace of $H_s$ corresponding to the zero eigenvalue 
is two-dimensional. An orthonormal basis can be constructed from 
the saddle-point state ${\mbox{\boldmath$\eta$}}_s(x)$. 
The translational invariance of $F$ gives 
$\nabla{\mbox{\boldmath$\eta$}}_s(x)$ as an eigenvector of $H_s$:
$H_s\nabla{\mbox{\boldmath$\eta$}}_s(x)={\bf 0}$. The gauge invariance
gives ${\bf u}_s^{(2)}(x)$ in Eq. (\ref{us2}) as another eigenvector 
of $H_s$: $H_s{\bf u}_s^{(2)}(x)={\bf 0}$. Both 
$\nabla{\mbox{\boldmath$\eta$}}_s(x)$ and ${\bf u}_s^{(2)}(x)$
are readily computed from ${\mbox{\boldmath$\eta$}}_s(x)$ numerically.
Based on these two nonorthogonal eigenvectors, the eigenvector
${\bf u}_s^{(3)}(x)$, normalized and orthogonal to ${\bf u}_s^{(2)}(x)$,
is obtained:
\begin{equation}\label{us3}
{\bf u}_s^{(3)}(x)=\alpha_s^{(3)}\left[
\nabla{\mbox{\boldmath$\eta$}}_s(x)-\left\langle
{\bf u}_s^{(2)}(x),\nabla{\mbox{\boldmath$\eta$}}_s(x)\right\rangle
{\bf u}_s^{(2)}(x)\right],
\end{equation}
where $\alpha_s^{(3)}$ is the normalization factor and
$\langle{\bf u}_s^{(2)}(x),\nabla{\mbox{\boldmath$\eta$}}_s(x)\rangle$ 
stands for the inner product 
$\int dx [{\bf u}_s^{(2)}(x)]^T\nabla{\mbox{\boldmath$\eta$}}(x)$.
Consider an infinitesimal variation of ${\mbox{\boldmath$\eta$}}_s(x)$,
$d{\mbox{\boldmath$\eta$}}_s(x)=\nabla{\mbox{\boldmath$\eta$}}_s(x)dx_c$,
where $dx_c$ denotes an infinitesimal translation of the phase-slip center.
The corresponding change of the coordinate in the direction of
${\bf u}_s^{(3)}(x)$ is given by
\begin{equation}\label{dcs3}
dc_s^{(3)}=\left\langle
{\bf u}_s^{(3)}(x),d{\mbox{\boldmath$\eta$}}_s(x)\right\rangle
=\left\langle{\bf u}_s^{(3)}(x),\nabla{\mbox{\boldmath$\eta$}}_s(x)
\right\rangle dx_c=\Lambda dx_c,
\end{equation}
where $\Lambda$ is related to the normalization factor $\alpha_s^{(3)}$
in Eq. (\ref{us3}) by $\Lambda=[\alpha_s^{(3)}]^{-1}$. Heuristically,
Eq. (\ref{us3}) defines ${\bf u}_s^{(3)}(x)$ to represent 
a special direction in the ${\mbox{\boldmath$\eta$}}$-function space.
Along this direction, the change of ${\mbox{\boldmath$\eta$}}_s(x)$
is a ``pure'' translation of the phase-slip center, without any rotation
of the global phase angle, which is an unphysical degree of freedom.
Then in Eq. (\ref{dcs3}), the projection of $d{\mbox{\boldmath$\eta$}}_s(x)$ 
onto ${\bf u}_s^{(3)}(x)$ measures the infinitesimal translation of 
the phase-slip center in the ${\mbox{\boldmath$\eta$}}$-function space,
along the physically nontrivial direction of ${\bf u}_s^{(3)}(x)$. 
With the help of Eq. (\ref{dcs3}) and $\int dx_c=l$, Eq. (\ref{zero-3})  
is obtained from Eq. (\ref{zero-2}).

To summarize, the dimensionless expressions for the prefactors 
$\Omega_{\pm}$ are obtained as
\begin{equation}\label{real-omega}
\begin{array}{ll}
\Omega_{+}=\displaystyle\frac{|\lambda_s^{(1)}|}{4\pi^2}
\sqrt{\displaystyle\frac{\sigma\xi H_c^2(T)}{k_BT}}\Lambda l
\left[\displaystyle\frac{\det' H_{n-1}}{|\det'' H_s|}
\right]^{1/2},\\
\Omega_{-}=\displaystyle\frac{|\lambda_s^{(1)}|}{4\pi^2}
\sqrt{\displaystyle\frac{\sigma\xi H_c^2(T)}{k_BT}}\Lambda l
\left[\displaystyle\frac{\det' H_n}{|\det'' H_s|}
\right]^{1/2},
\end{array}
\end{equation}
where the zero eigenvalues $\lambda_n^{(1)}$ and $\lambda_s^{(2)}$ are 
omitted and Eq. (\ref{zero-3}) is used for $[\lambda_s^{(3)}]^{-1/2}$.
Here $\det'$ in $\det' H_{n-1}$ and $\det' H_n$ indicates that
the only zero eigenvalue is to be omitted when computing 
the determinant, and $\det''$ in $\det'' H_s$ indicates that
the two zero eigenvalues are to be omitted when computing 
the determinant. Using those relevant eigenvectors 
(corresponding to the negative and zero eigenvalues),
the matrices $H_{n-1}$, $H_n$, and $H_s$
can all be modified into positive definite matrices
(as expressed by Eq. (\ref{positive-definite})),  for which
the determinant ratio can be computed according to 
Eq. (\ref{determinant-ratio}).

\section{Numerical Results}\label{numerical}
\subsection{Minimal Energy Path}\label{mep}
The string method has been employed to calculate the MEPs 
connecting neighboring metastable states $\psi_n$ and $\psi_{n-1}$. 
All quantities in the numerical calculation are dimensionless. 
The length of the system is $l=32\pi$. The local minima of $F$ 
are given in Eq. (\ref{metastable}) with $|n|\le 9$.
(The maximum $|n|$ allowed by $|k_n|<k_c$ is $9$.)

\begin{figure}
\centerline{\psfig{figure=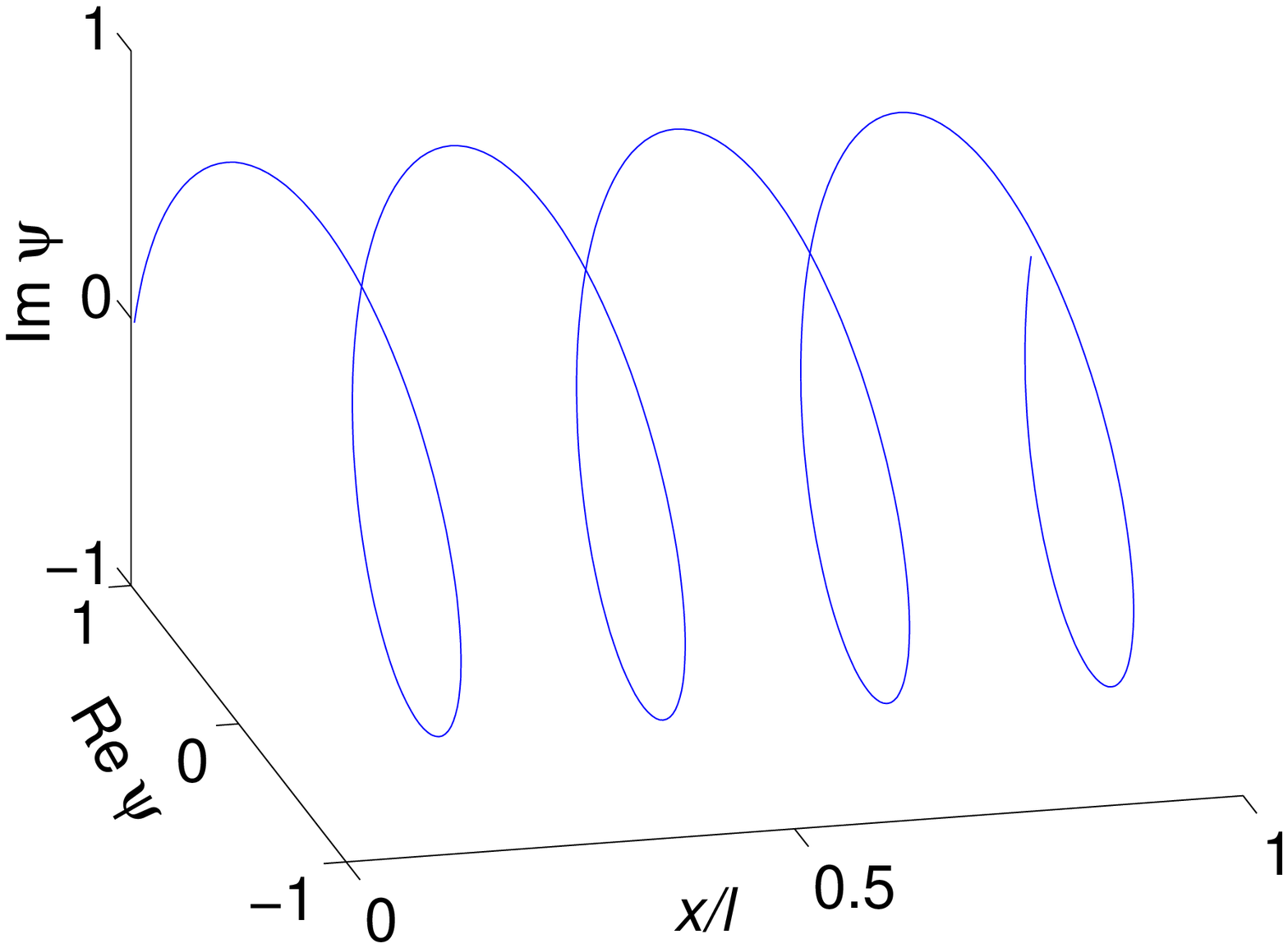,height=2.6cm} \hspace{1.5cm}
 \psfig{figure=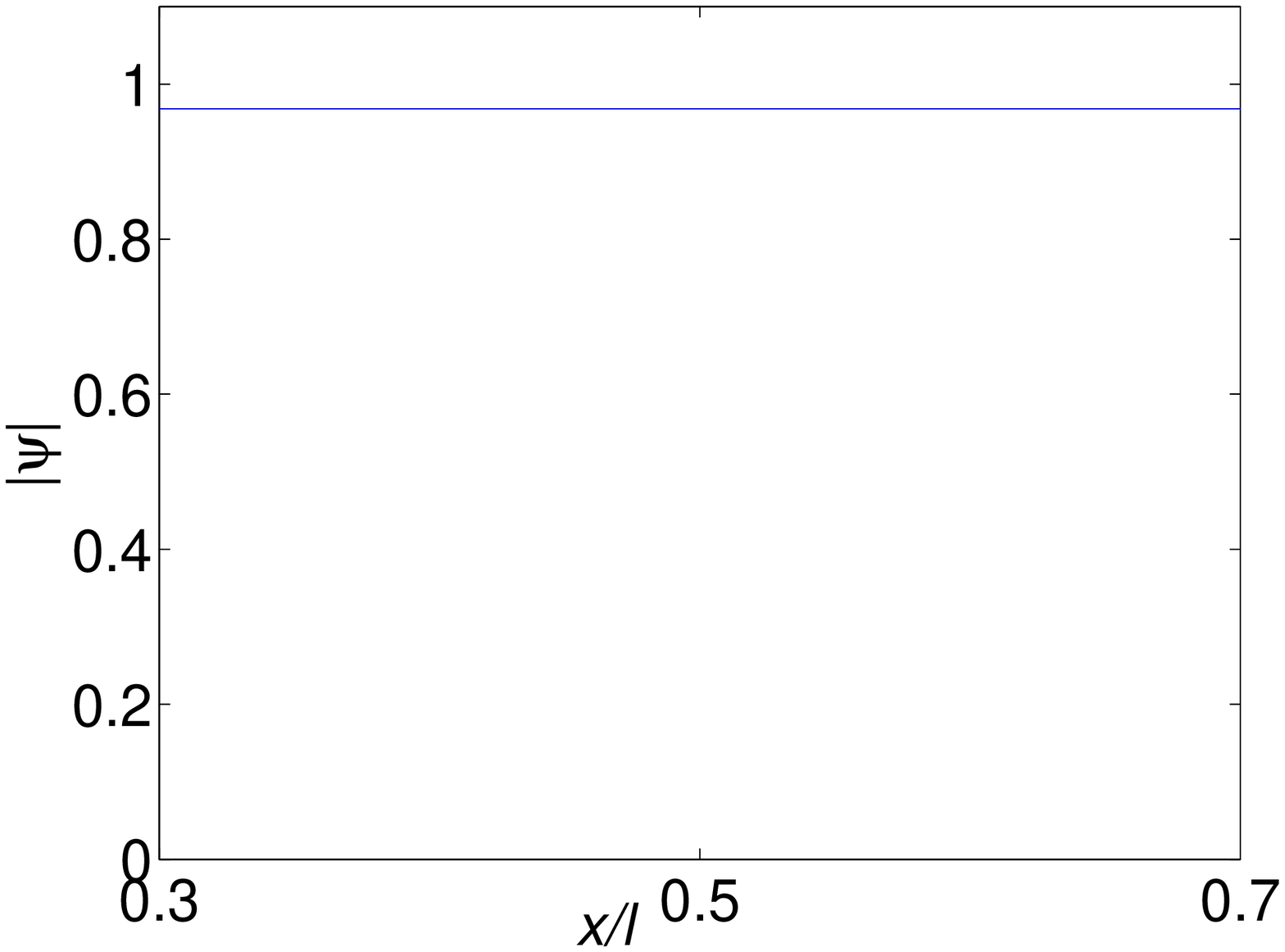,height=2.6cm}}
\centerline{\psfig{figure=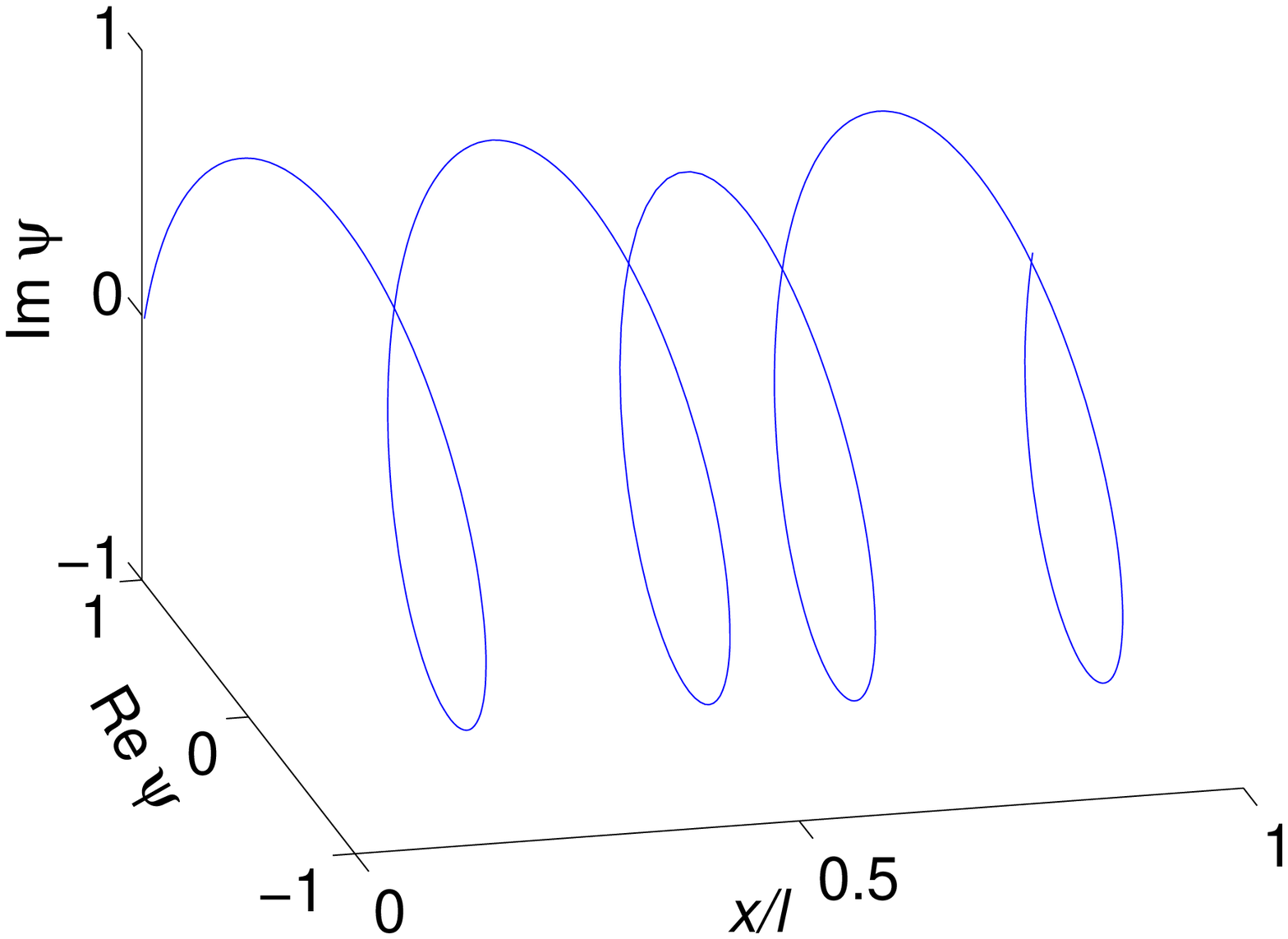,height=2.6cm} \hspace{1.5cm}
 \psfig{figure=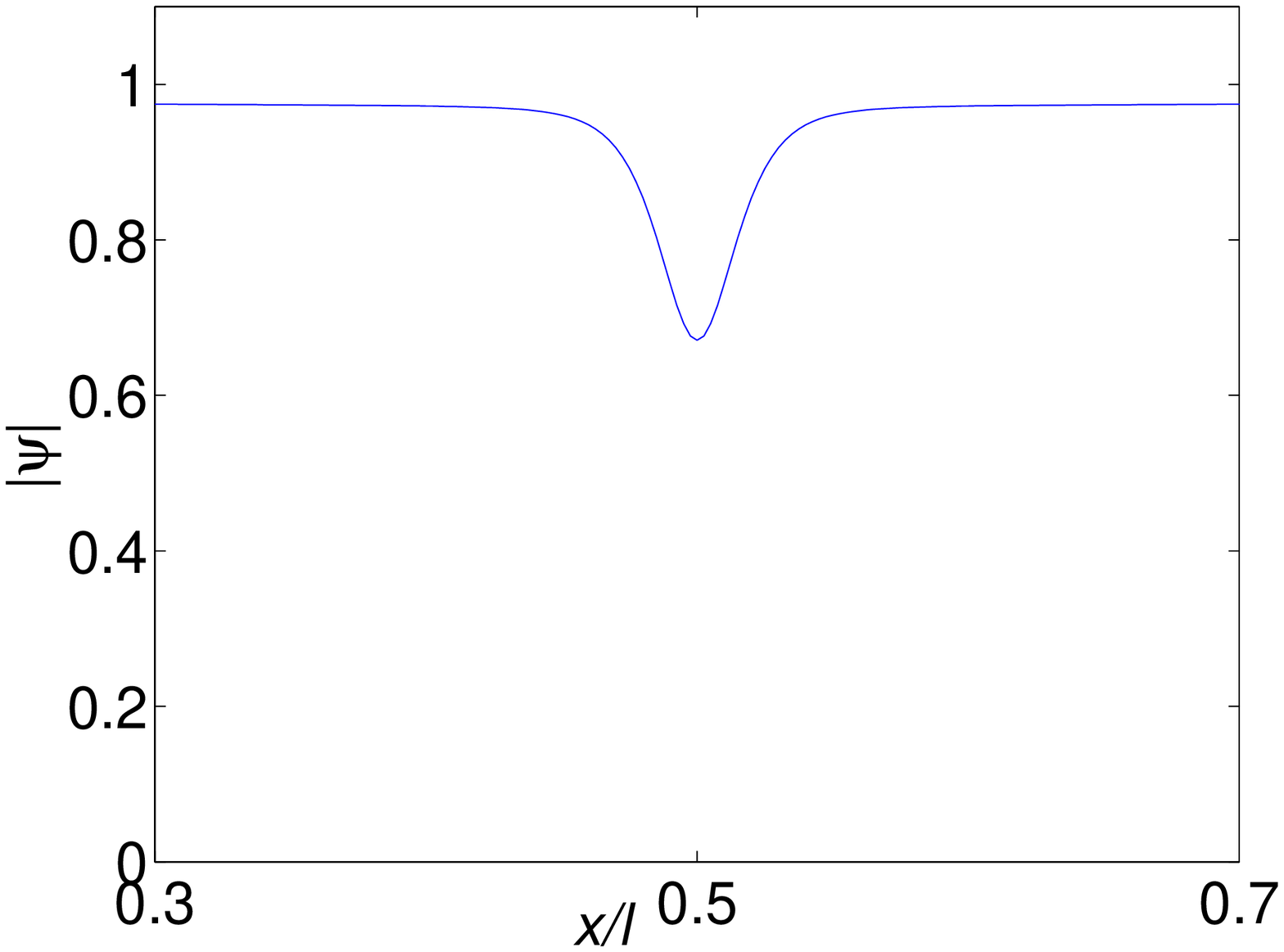,height=2.6cm}}
\centerline{\psfig{figure=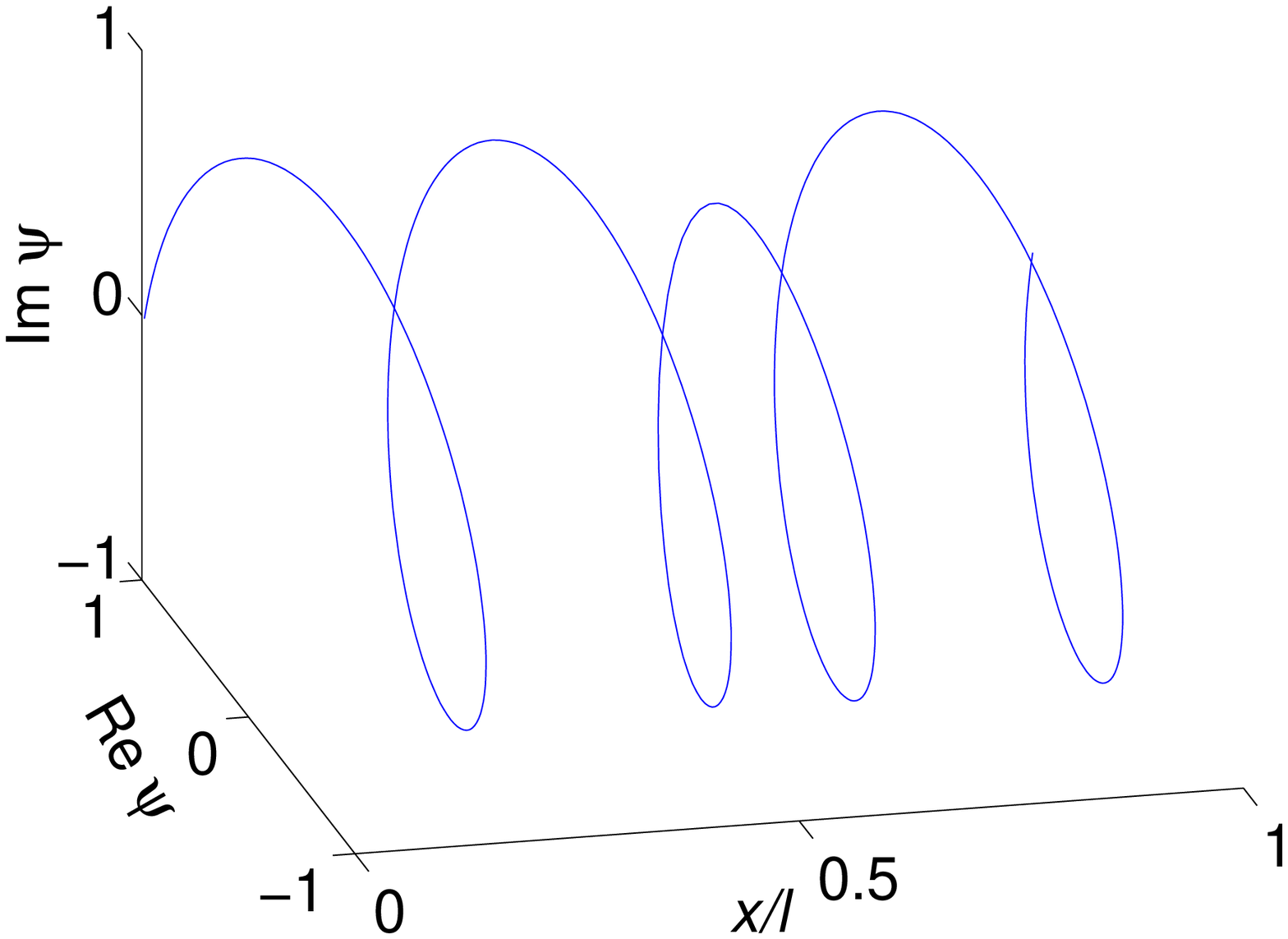,height=2.6cm} \hspace{1.5cm}
 \psfig{figure=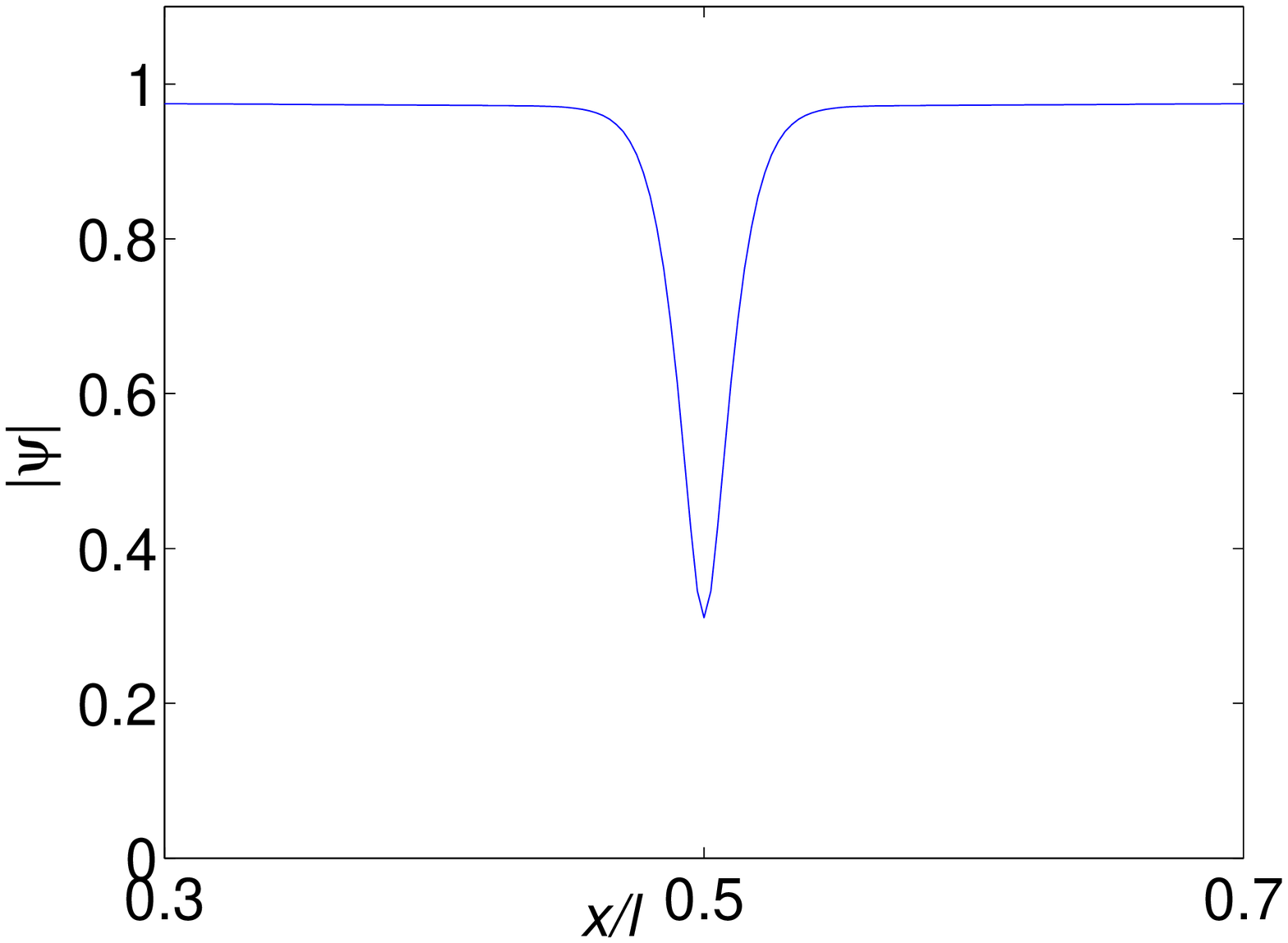,height=2.6cm}}
\centerline{\psfig{figure=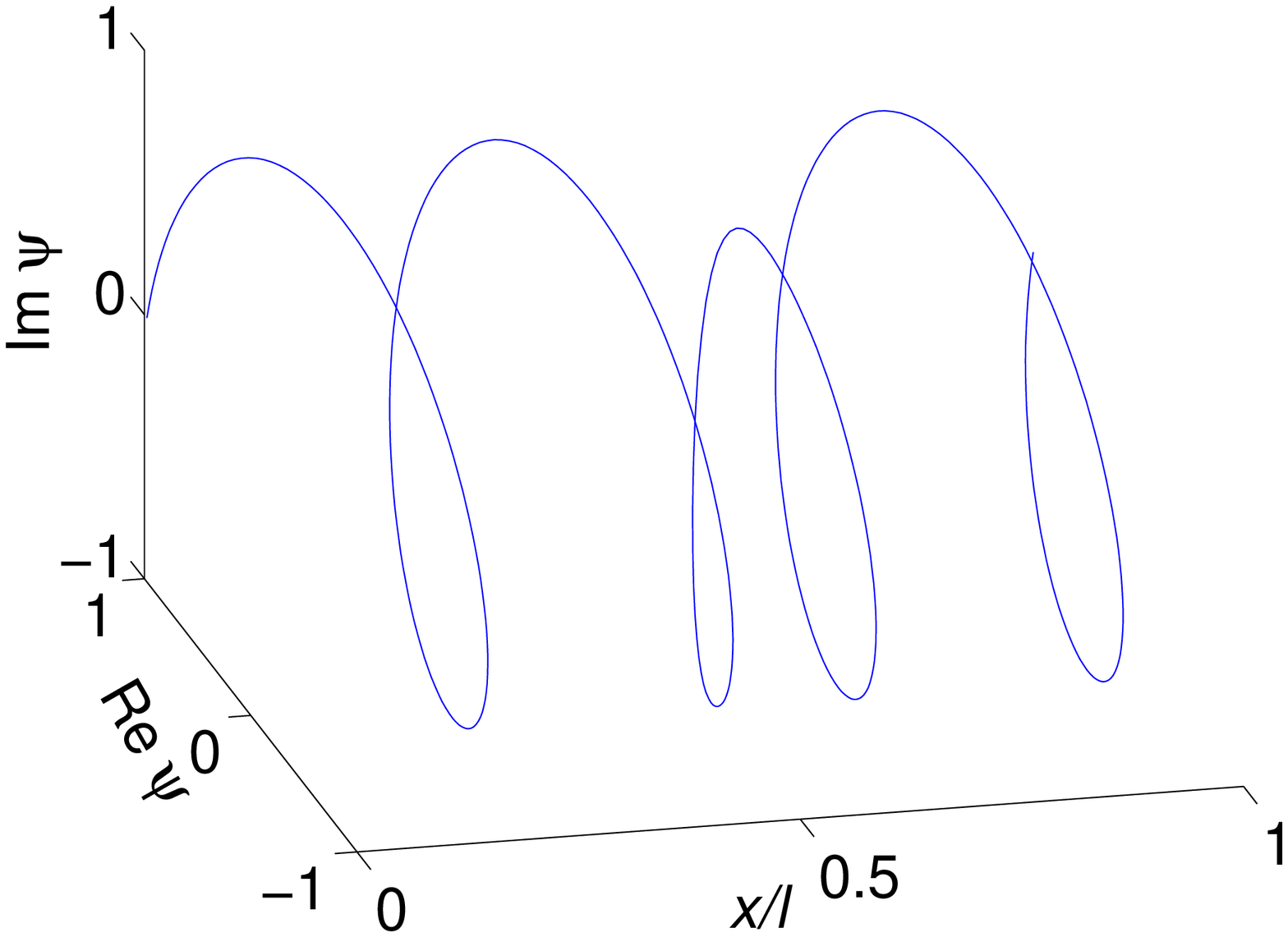,height=2.6cm}  \hspace{1.5cm}
 \psfig{figure=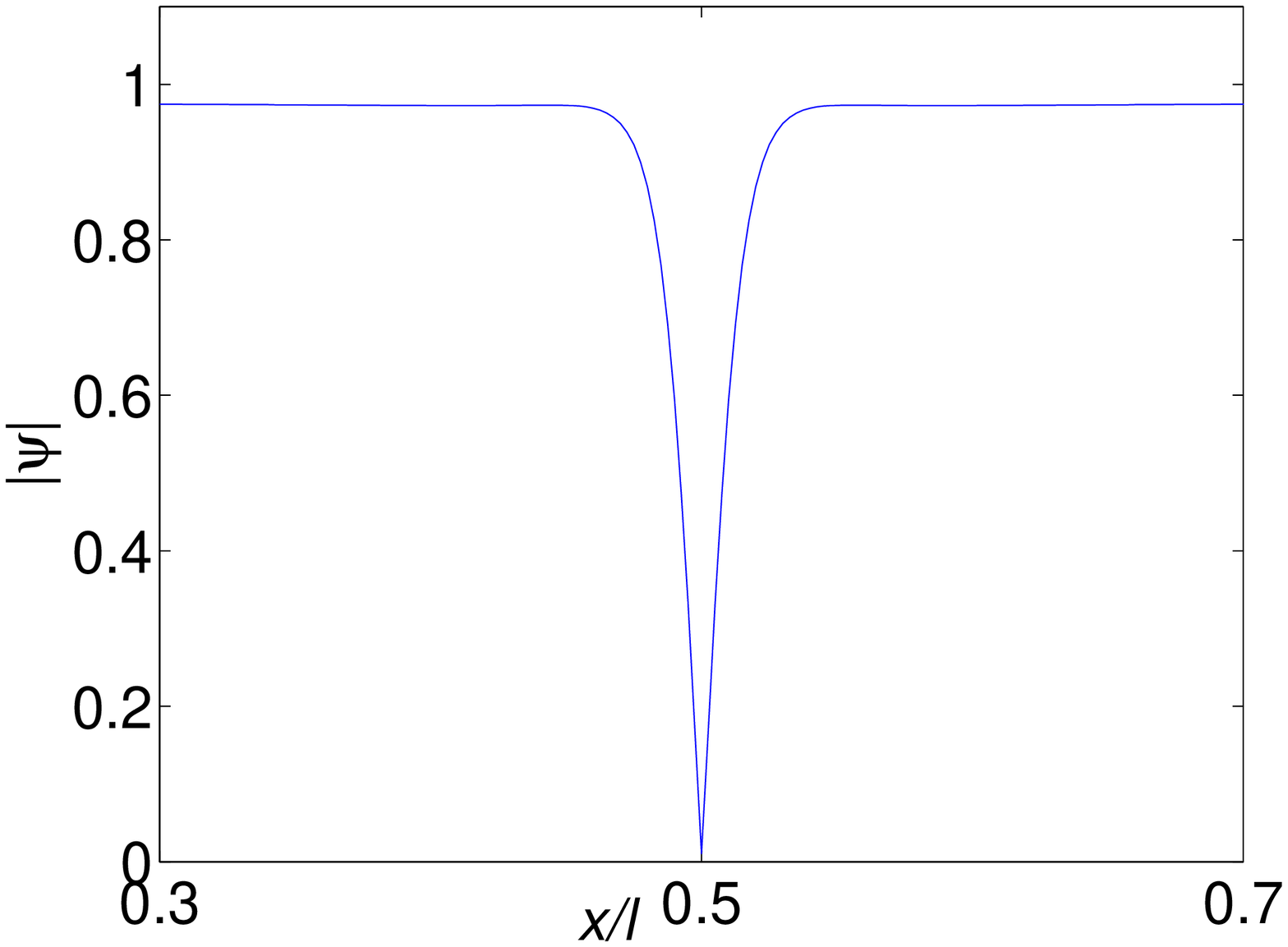,height=2.6cm}}
\centerline{\psfig{figure=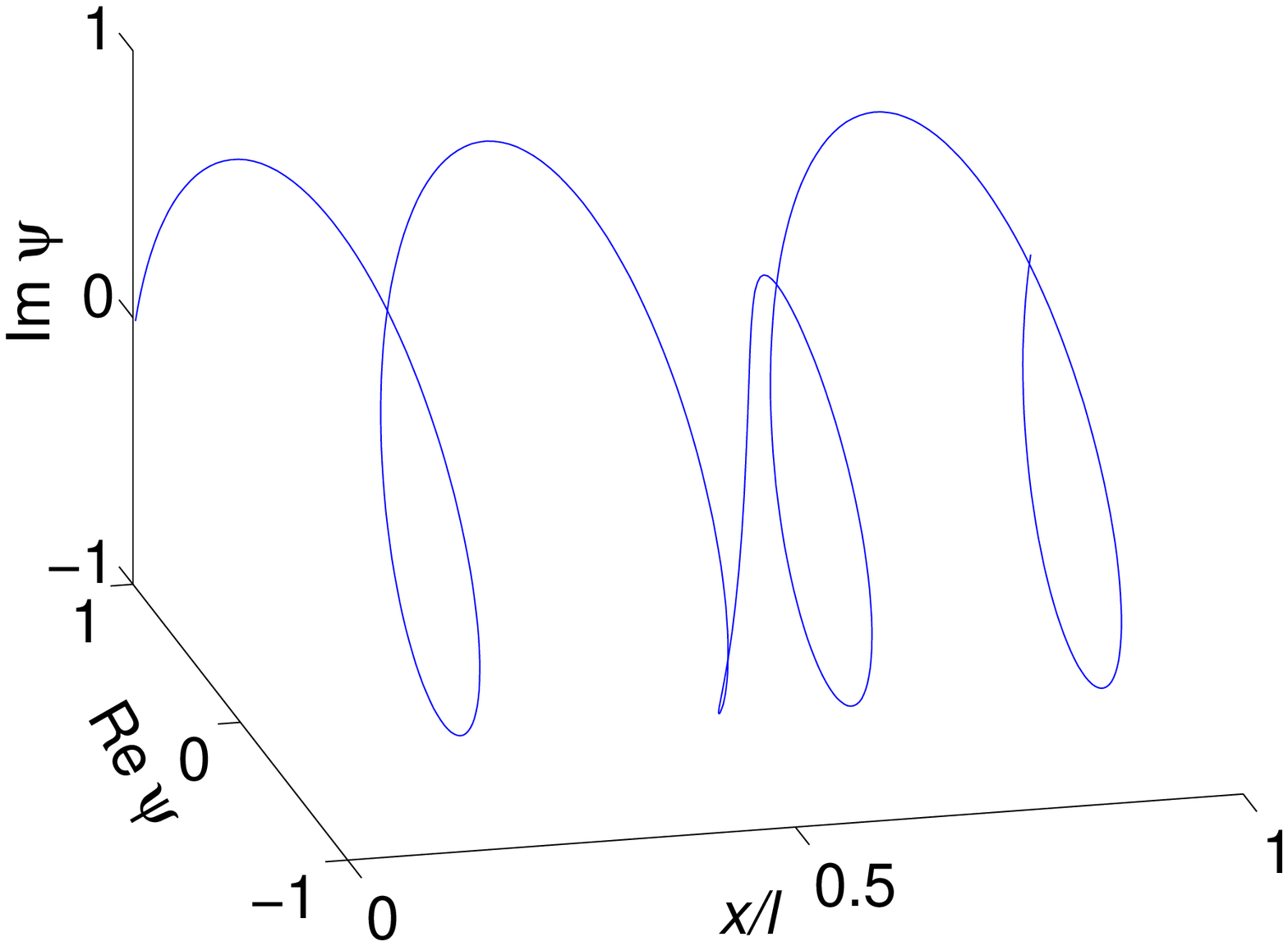,height=2.6cm} \hspace{1.5cm}
 \psfig{figure=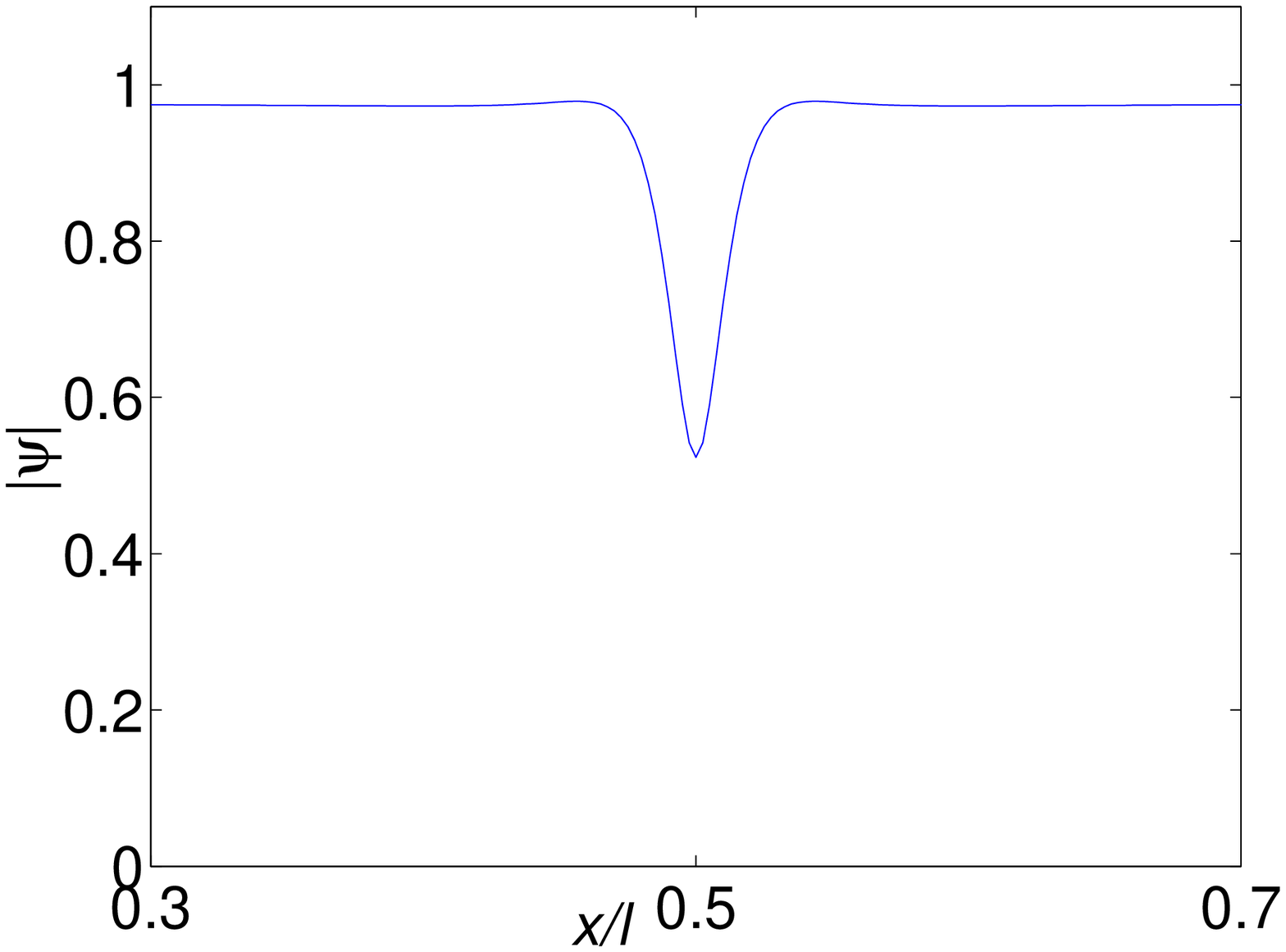,height=2.6cm}}
\centerline{\psfig{figure=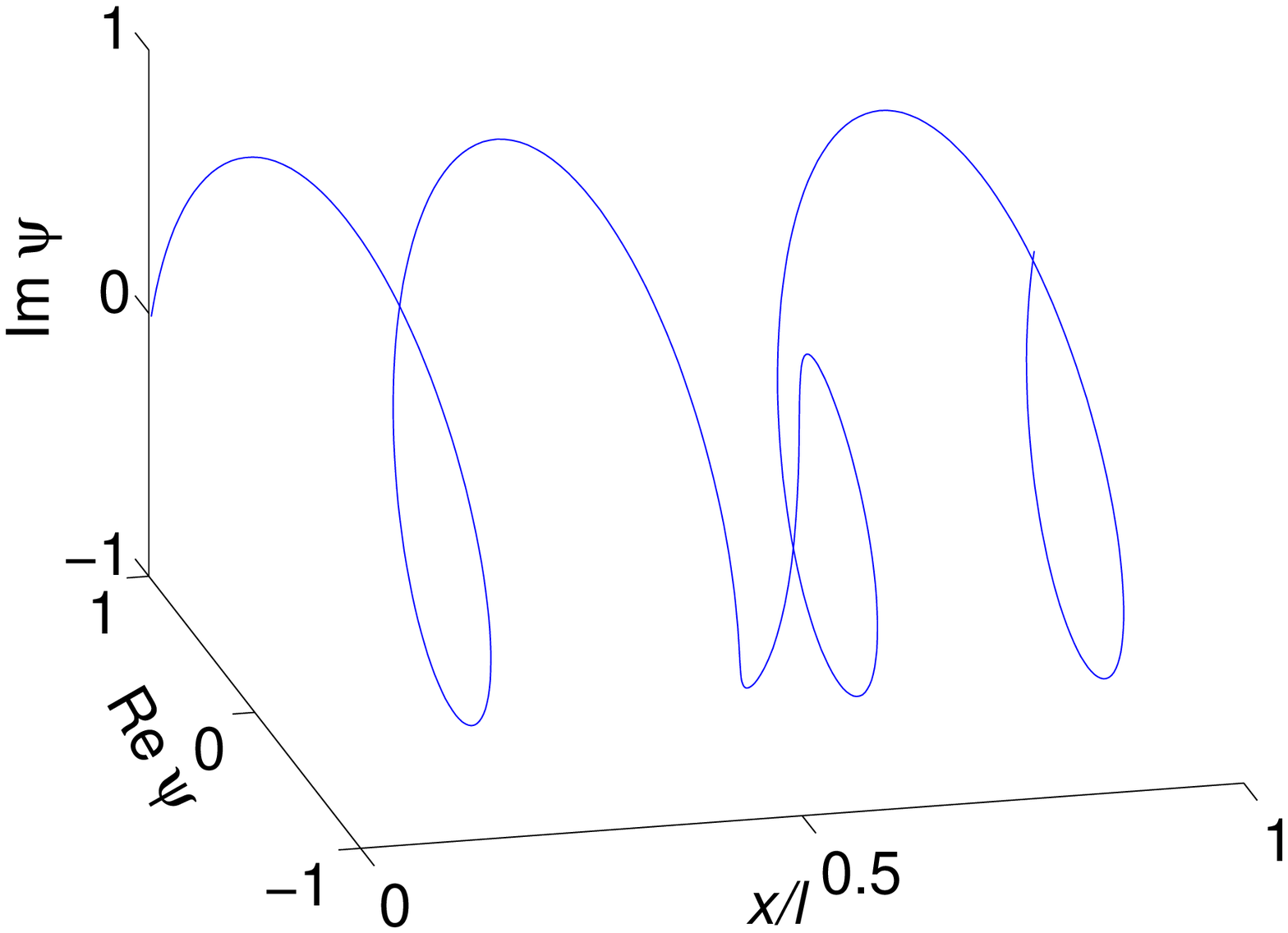,height=2.6cm} \hspace{1.5cm}
 \psfig{figure=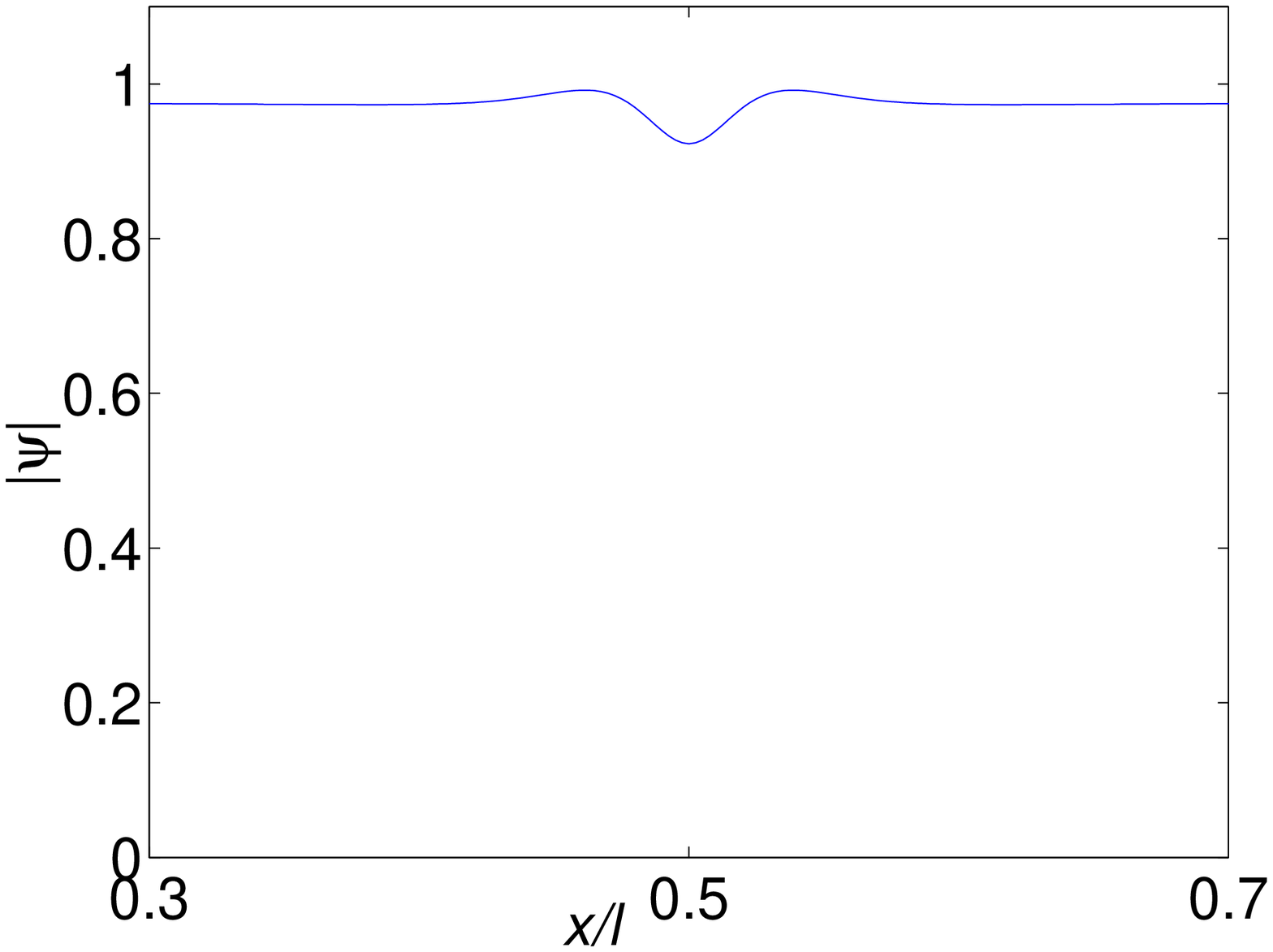,height=2.6cm}}
\centerline{\psfig{figure=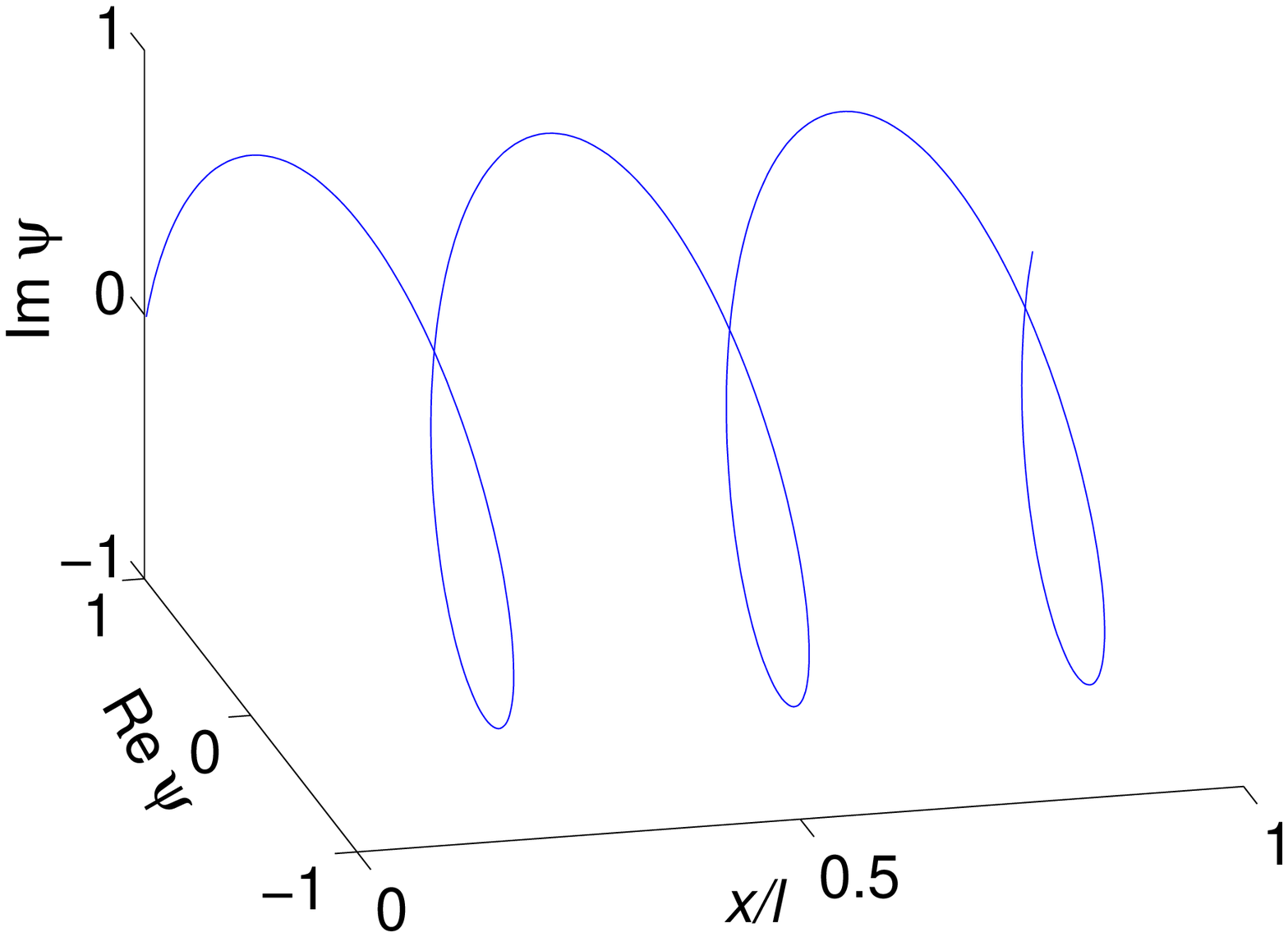,height=2.6cm} \hspace{1.5cm}
 \psfig{figure=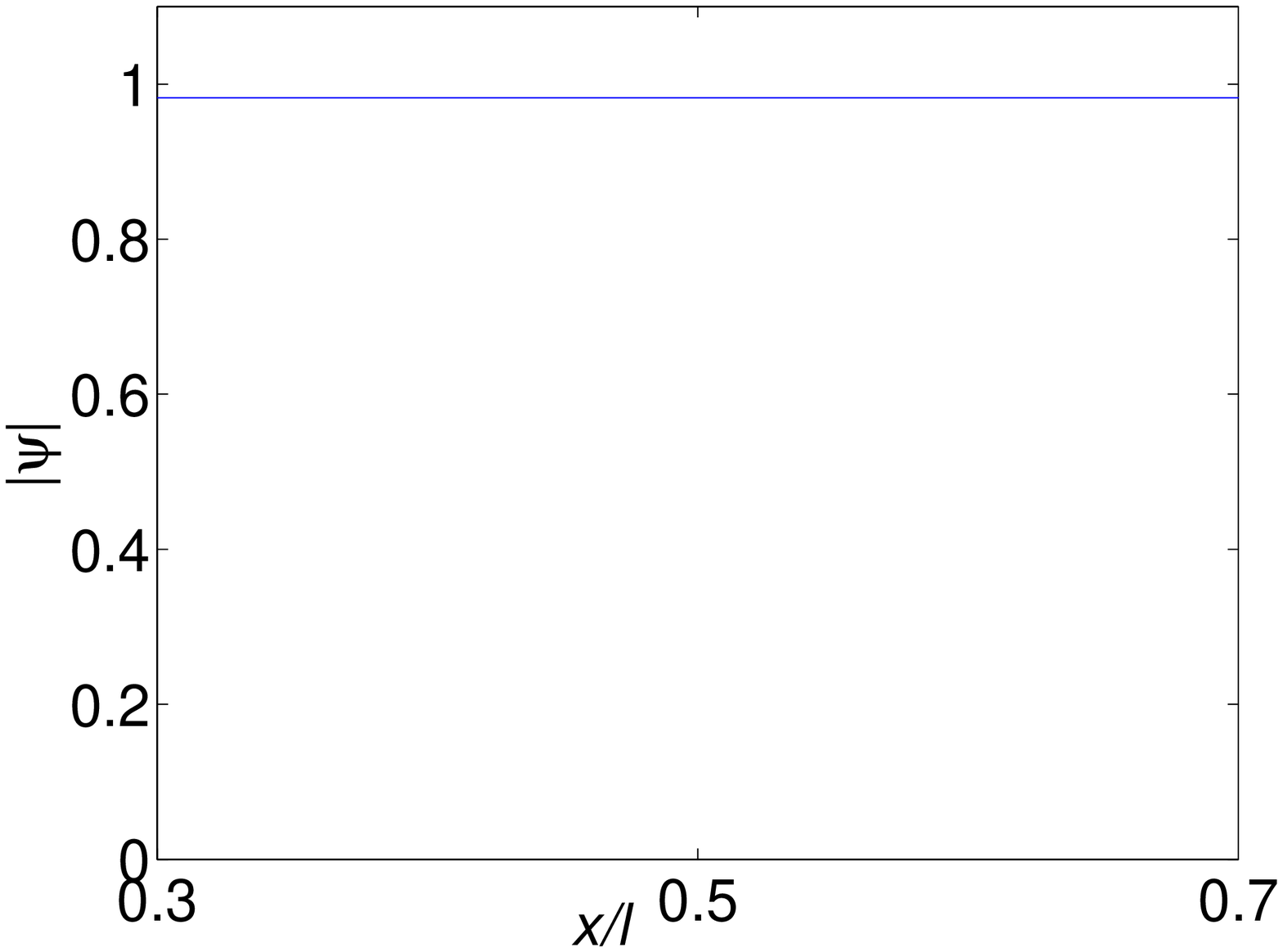,height=2.6cm}}
\bigskip
\caption{Minimal energy path from $\psi_4$ to $\psi_3$. 
Left column: $(Re \psi,Im \psi)$ as a function of $x$; 
Right column: $|\psi|$ as a function of $x$. 
The figures at the top and bottom correspond to $\psi_4$ and $\psi_3$, 
respectively. The third from the top is for the saddle point.
In the fourth figure, $|\psi|$ reaches zero in the middle.}\label{fig-MEP43}
\end{figure}

We first show in Fig \ref{fig-MEP43} the MEP which connects 
$\psi_4$ to $\psi_3$. The string is discretized by $M=101$ points
in the $\psi(x)$-function space. The initial string is taken from 
a linear interpolation between $\psi_4(x)$ and $\psi_3(x)$. 
In order to reach the MEP, the string is evolved toward 
the steady state according to Eq. (\ref{string-evolution}), 
with the potential force given by
$$
-\displaystyle\frac{\delta}{\delta {\mbox{\boldmath$\eta$}}}
F[{\mbox{\boldmath$\eta$}}(x)]=
\nabla^2{\mbox{\boldmath$\eta$}}+{\mbox{\boldmath$\eta$}}
-{\mbox{\boldmath$\eta$}}^2{\mbox{\boldmath$\eta$}}.
$$
During this process, the string is re-parametrized by arc length 
every 10 steps. In the calculation, $\psi(x)$ is represented 
by a column vector of $2N$ entries, with the $x$ interval $[0,l]$
discretized by a uniform mesh of $N=100$ points. 
Spatial derivatives in the potential force are discretized 
using central finite difference.

To fix the global rotation of the system, a spring force is applied to
the endpoint order parameter $\psi(0)$. In the form of 
$f_0=-{\cal K}\eta_2(0)$ with ${\cal K}=50$, this force restricts 
$\psi(0)$ to the real axis.
 
The first column in Fig \ref{fig-MEP43} displays a sequence of 
the configurations along the MEP from $\psi_4(x)$ to $\psi_3(x)$,
and the second column displays the corresponding sequence of
$|\psi(x)|$. Along this particular MEP, there is a phase slip of $2\pi$, 
nucleated in the middle of the system. Through this phase slip, 
the winding number changes from $n=4$ to $n-1=3$. 
From Fig. \ref{fig-MEP43}, it is seen that $|\psi(x)|$ first 
decreases and reaches zero somewhere (at $x=l/2$, 
see the fourth figure from the top), then the phase slip occurs 
and $|\psi(x)|$ rebounds to accomplish the transition.    
The third figure from the top shows the saddle point $\psi_s(x)$ between 
$\psi_4(x)$ and $\psi_3(x)$, which has a locally diminished amplitude
and possesses the highest energy along the MEP. 

Little \cite{little} first pointed out that a persistent current
in a closed loop will not be destroyed, ``unless a fluctuation occurs
which is of such an amplitude that the order parameter is driven to zero 
for some section of the loop''. However, the configuration of 
a vanishing order-parameter amplitude somewhere does not necessarily
correspond to the lowest saddle point between two current-carrying 
metastable states. Using the stationary Ginzburg-Landau equation, 
Langer and Ambegaokar \cite{LA} have obtained the analytical solution
for the free-energy saddle point. They also pointed out the following:
``It is plausible that, from this state of locally diminished amplitude, 
the system will run downhill in free energy through a configuration
in which the amplitude vanishes somewhere, and finally will achieve
the configuration in which one less loop in $\psi$ occurs across 
the length $L$.'' This picture about the transition pathway has been
quantitatively confirmed by the MEP obtained here.

\begin{figure}
\centerline{\psfig{figure=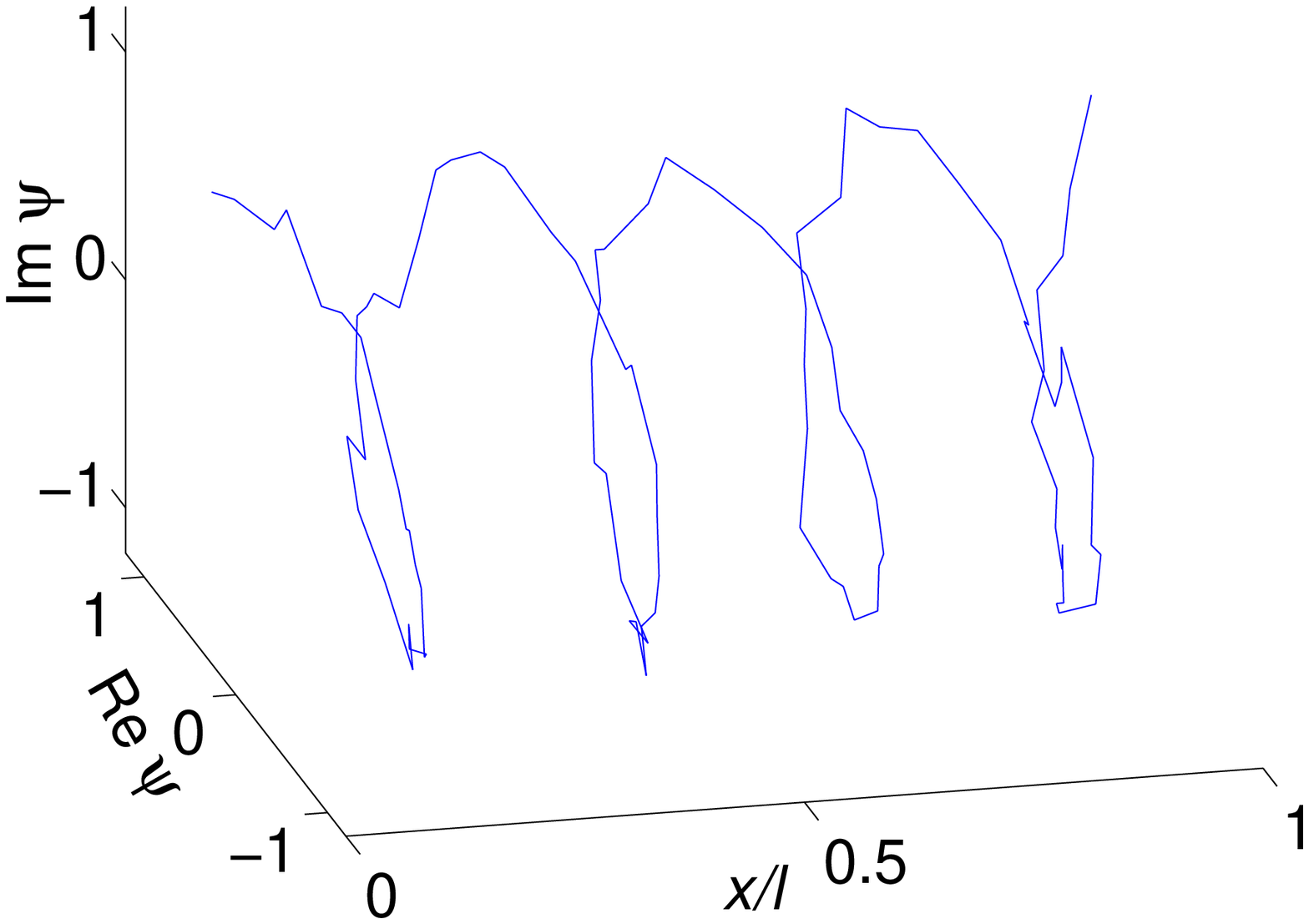,height=2.6cm} \hspace{1.5cm}
 \psfig{figure=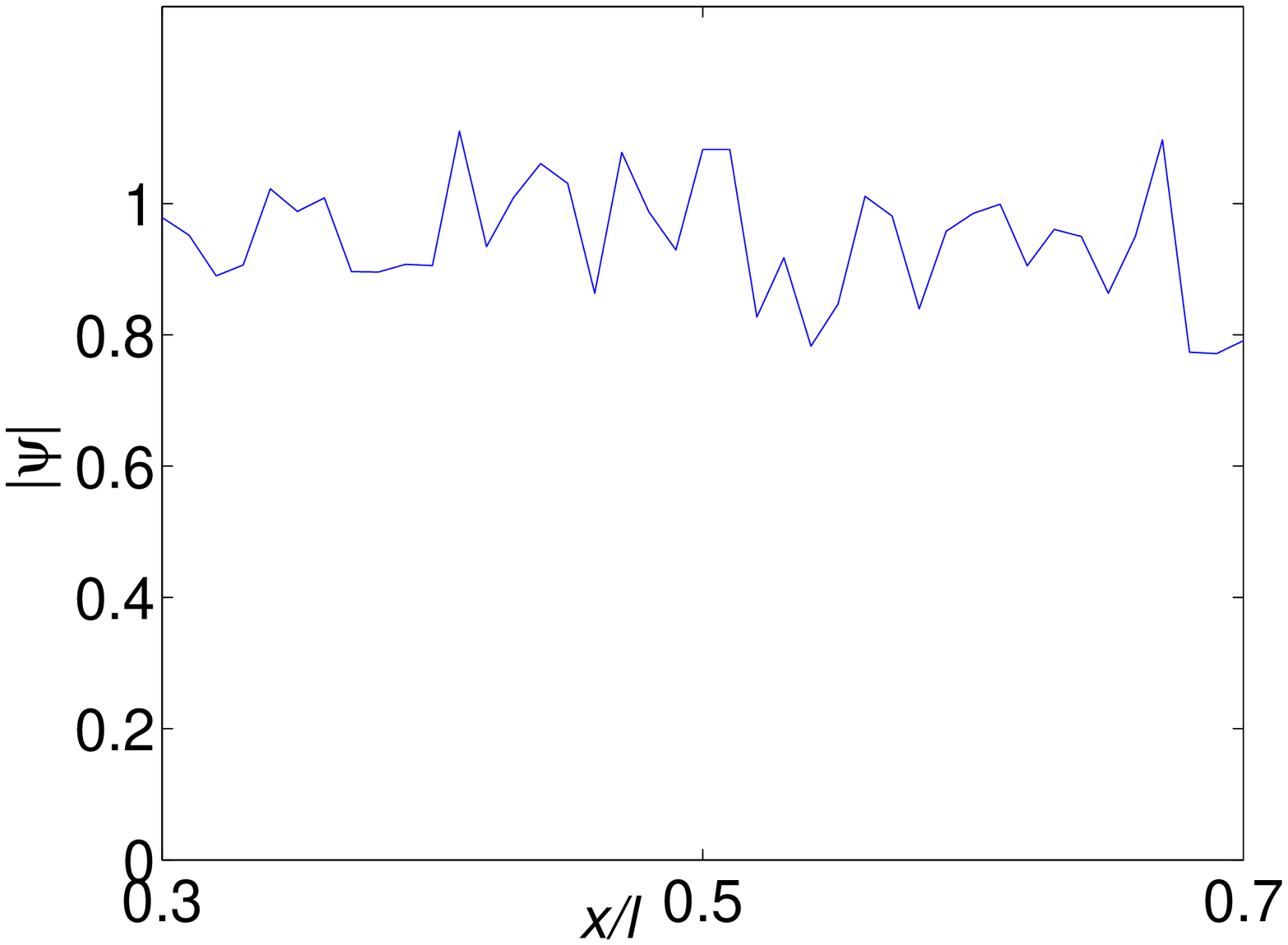,height=2.6cm}}
\centerline{\psfig{figure=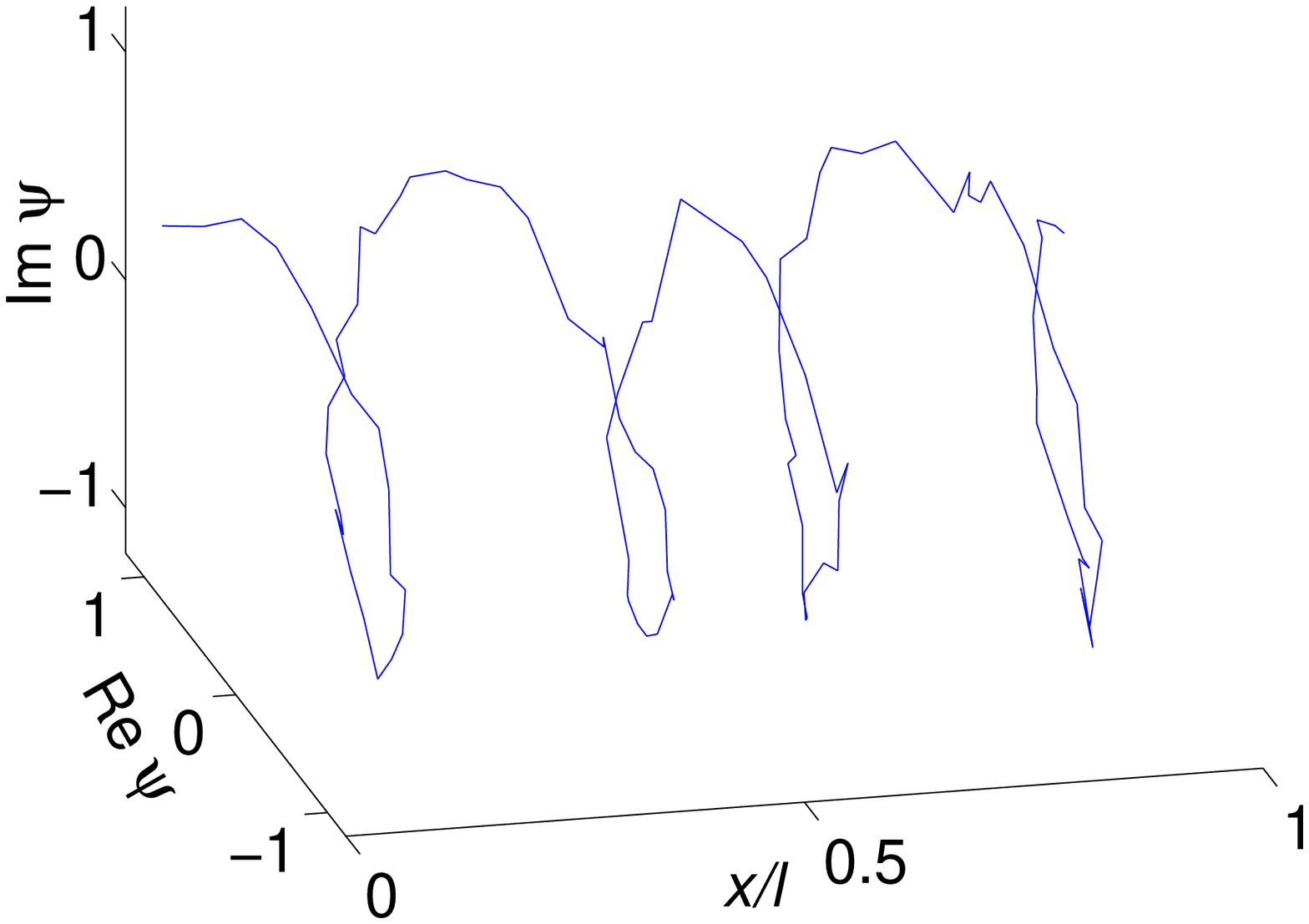,height=2.6cm} \hspace{1.5cm}
 \psfig{figure=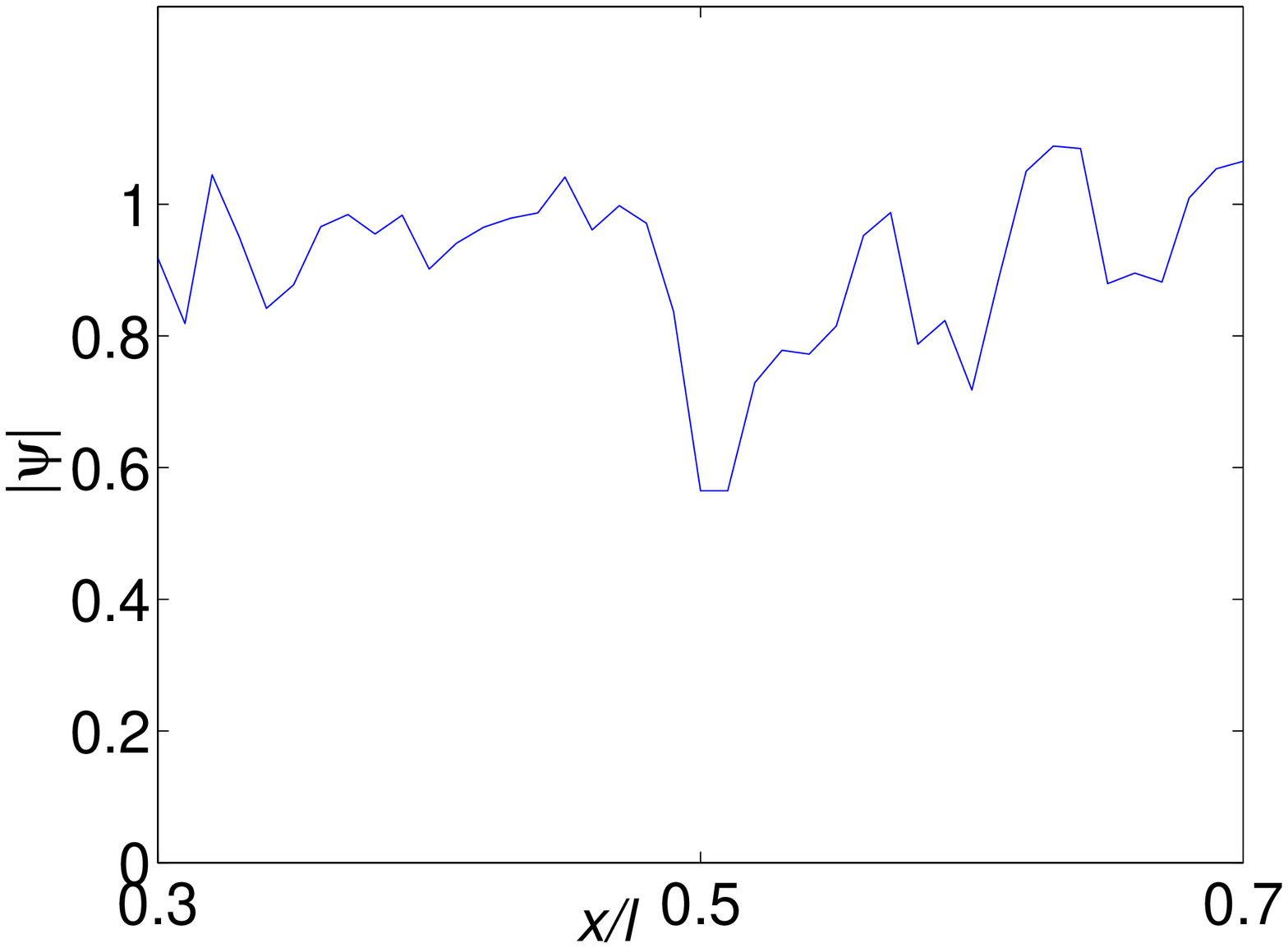,height=2.6cm}}
\centerline{\psfig{figure=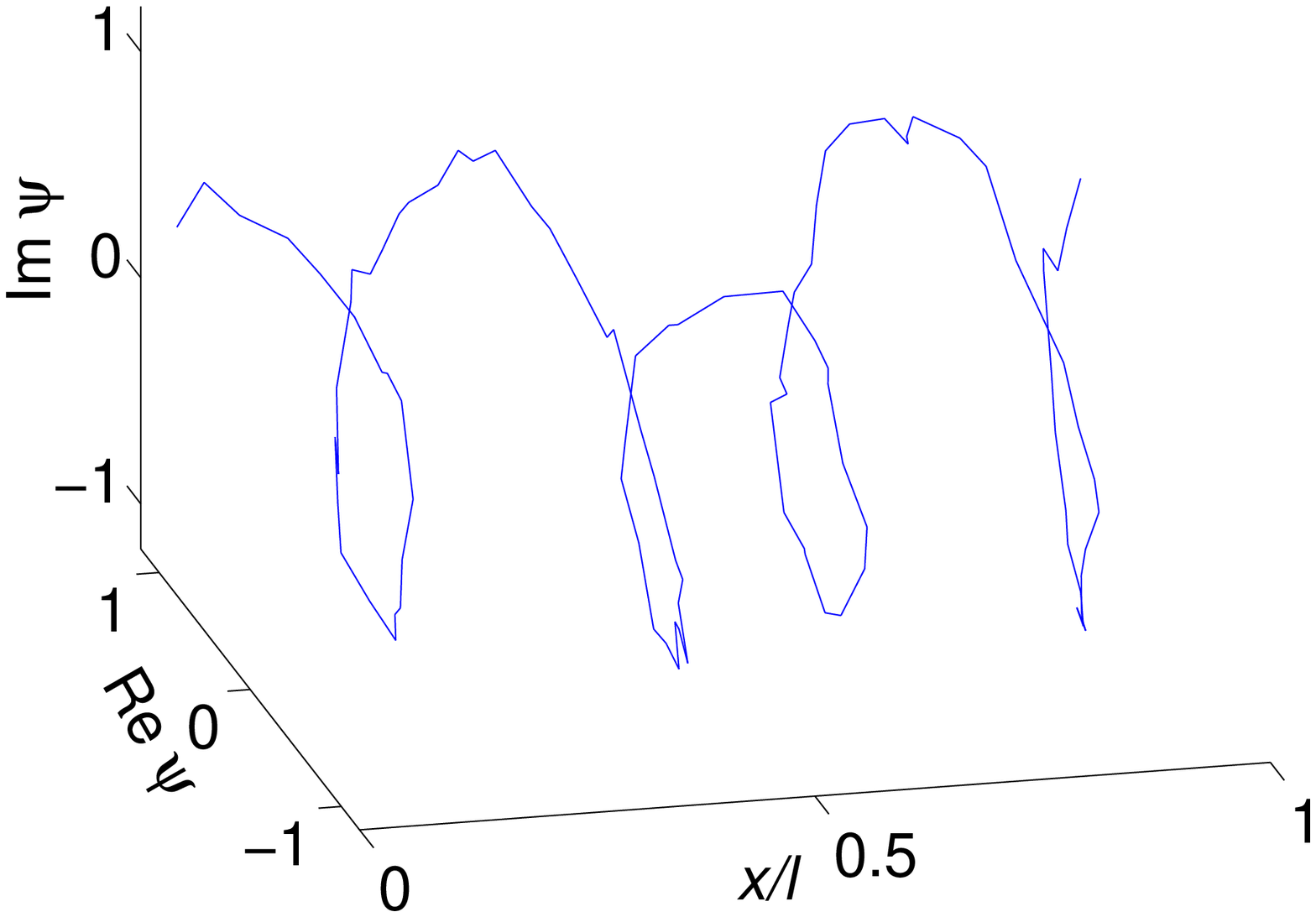,height=2.6cm} \hspace{1.5cm}
 \psfig{figure=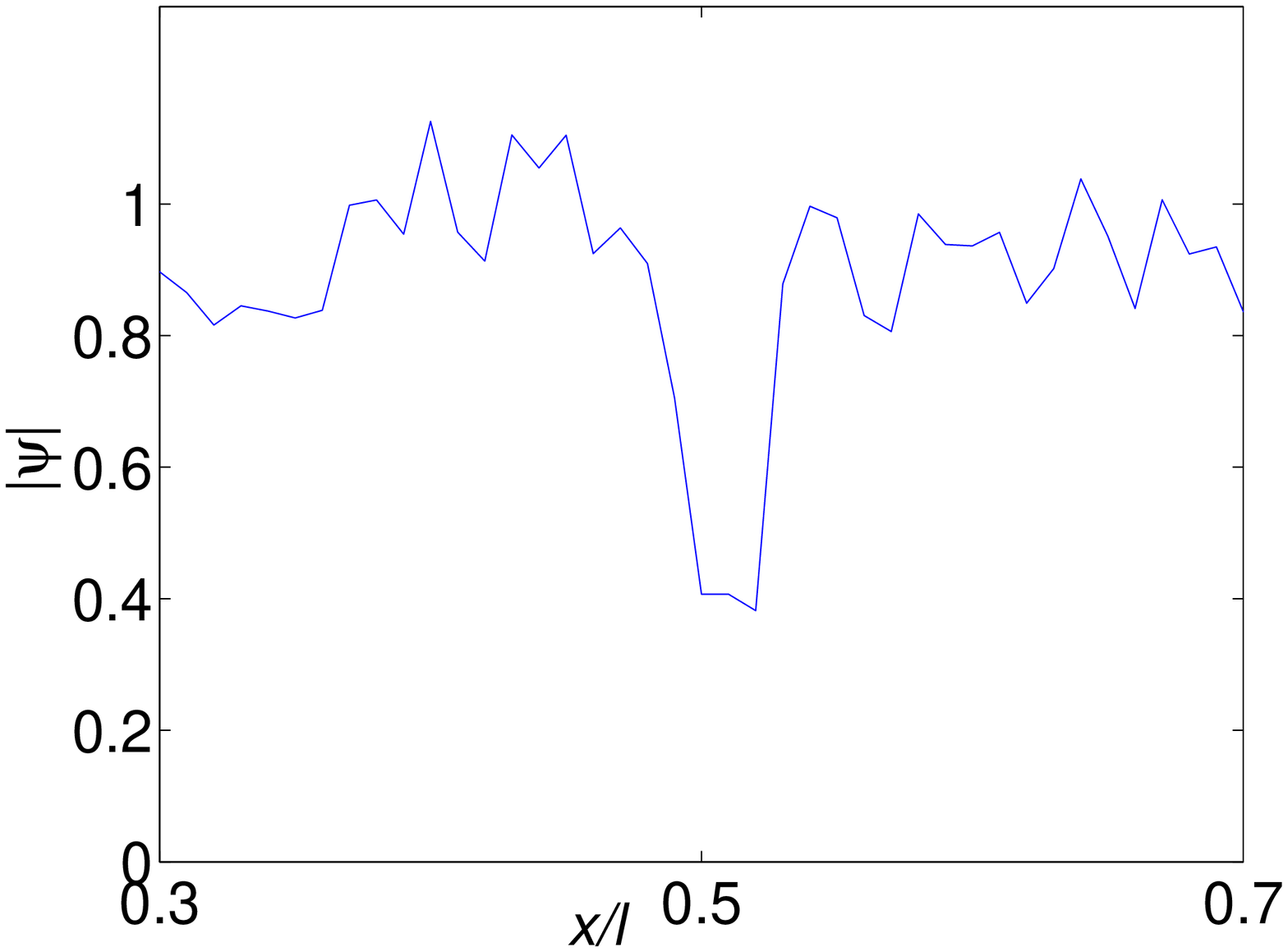,height=2.6cm}}
\centerline{\psfig{figure=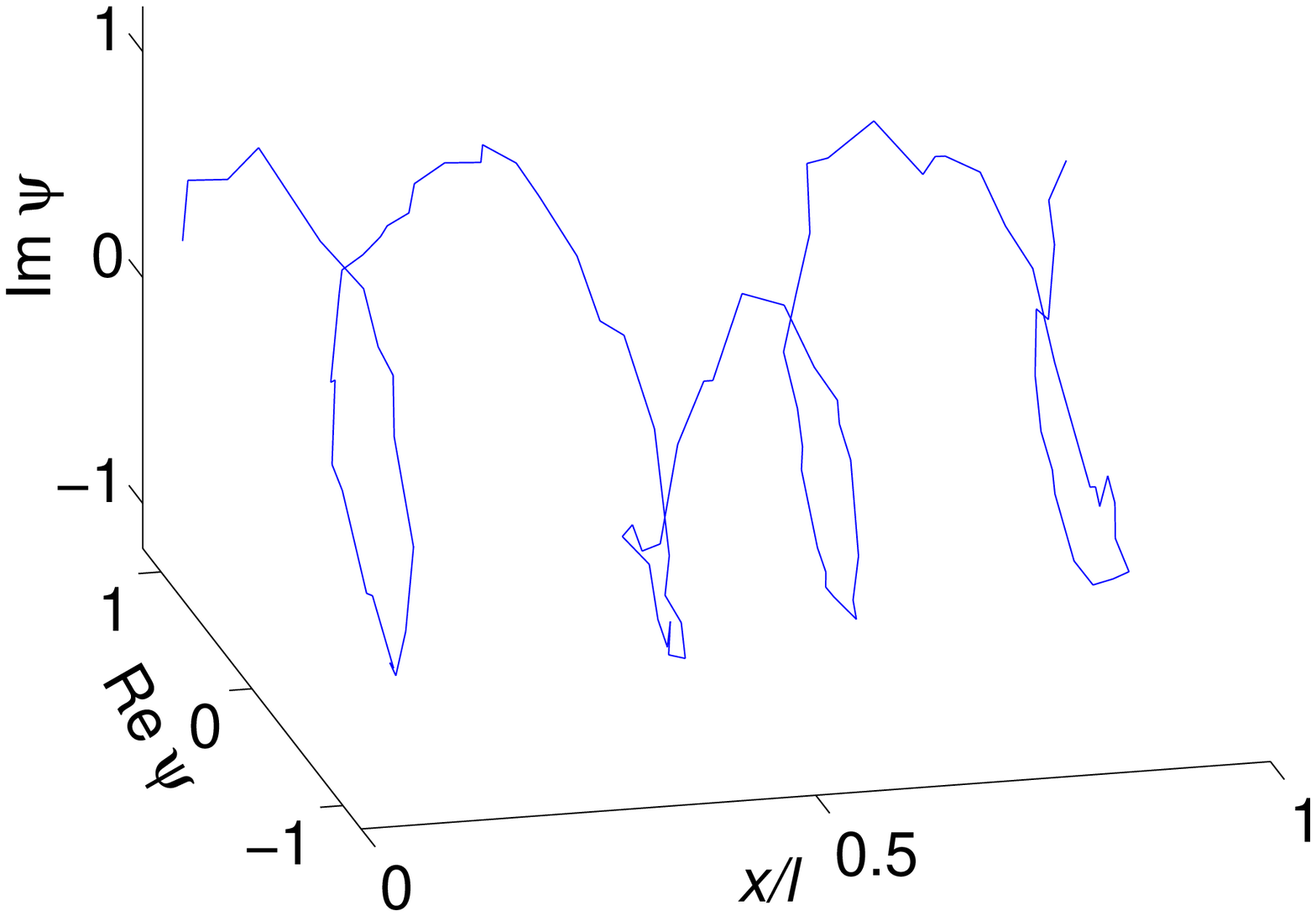,height=2.6cm}  \hspace{1.5cm}
 \psfig{figure=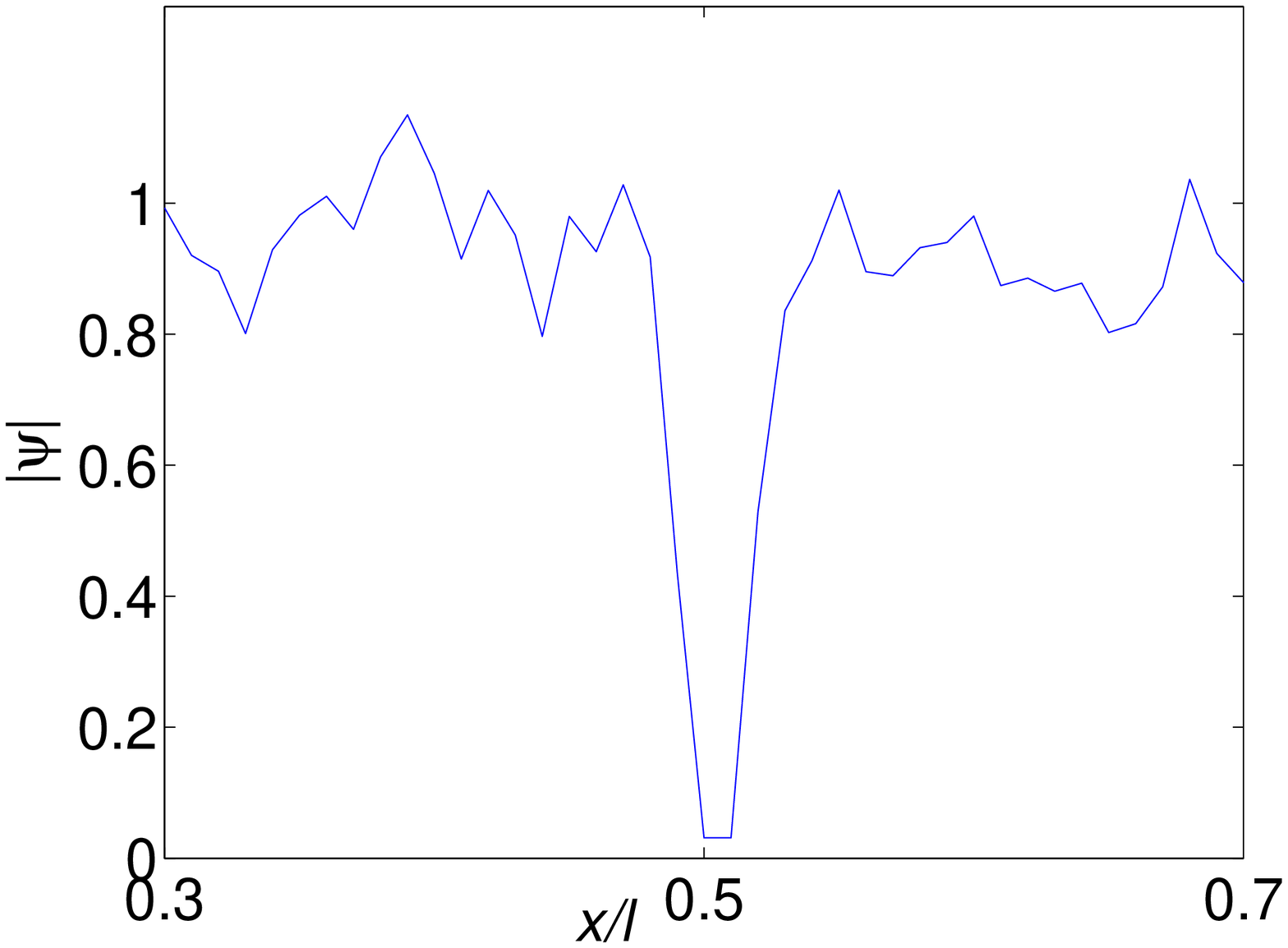,height=2.6cm}}
\centerline{\psfig{figure=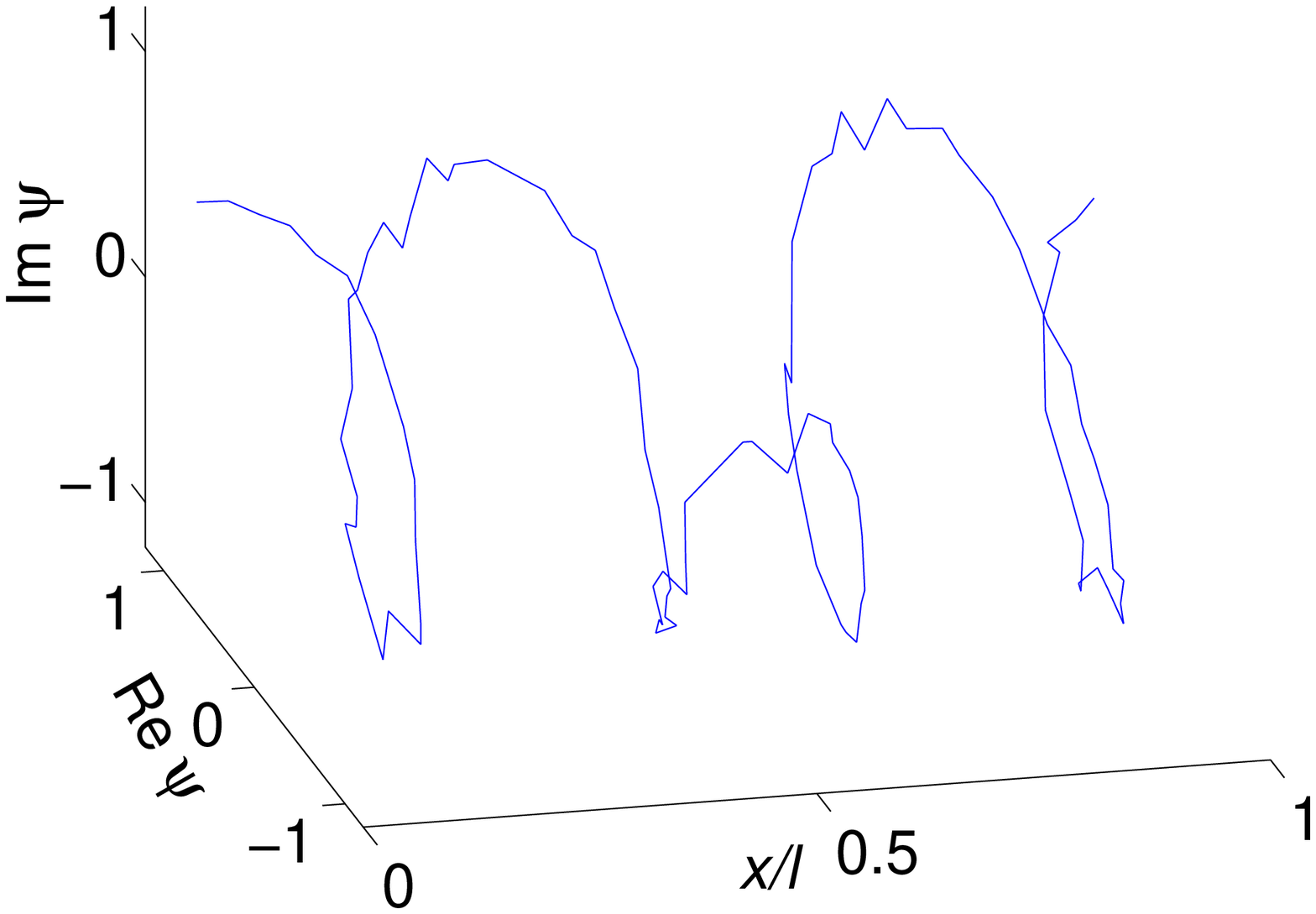,height=2.6cm} \hspace{1.5cm}
 \psfig{figure=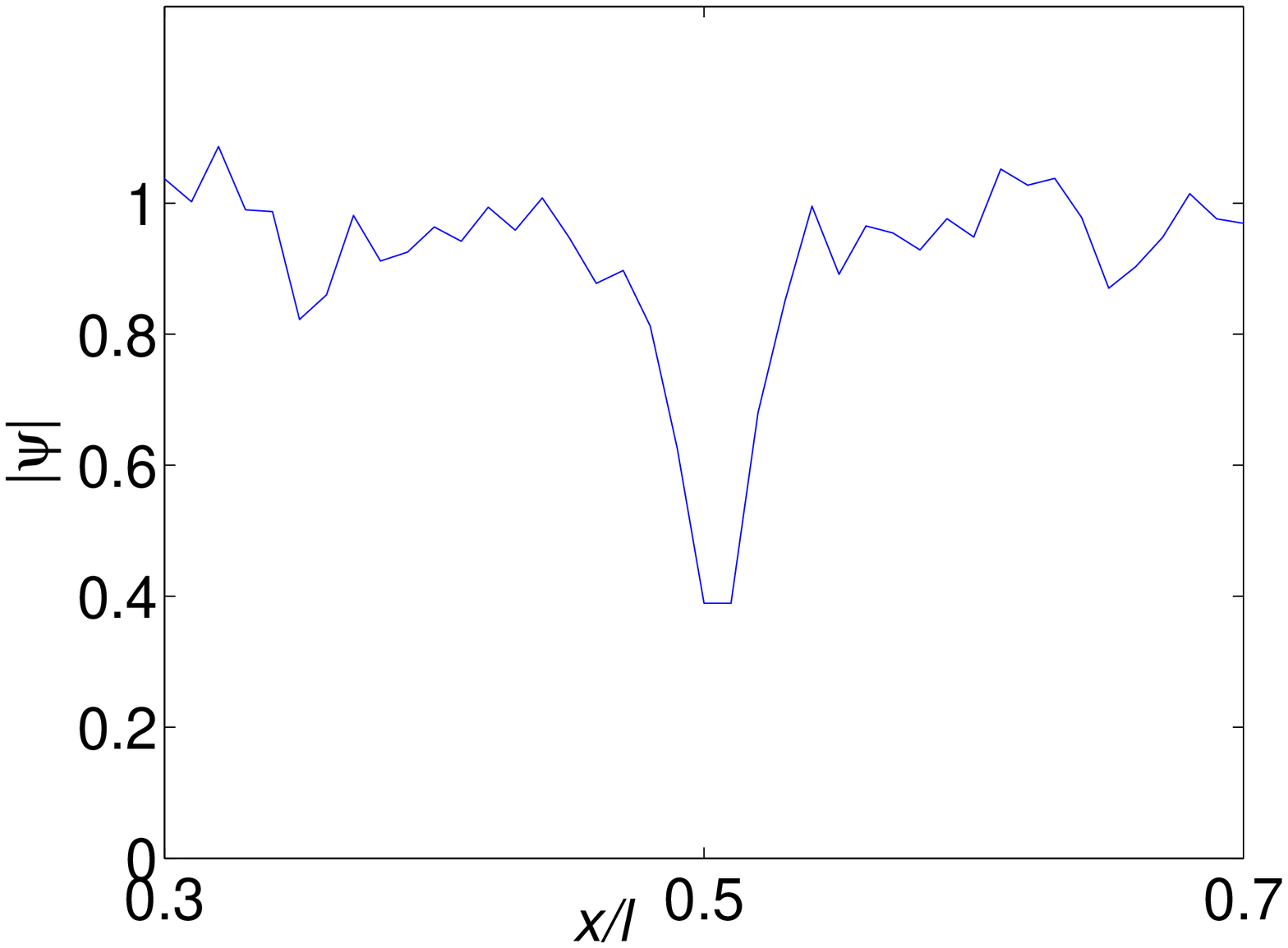,height=2.6cm}}
\centerline{\psfig{figure=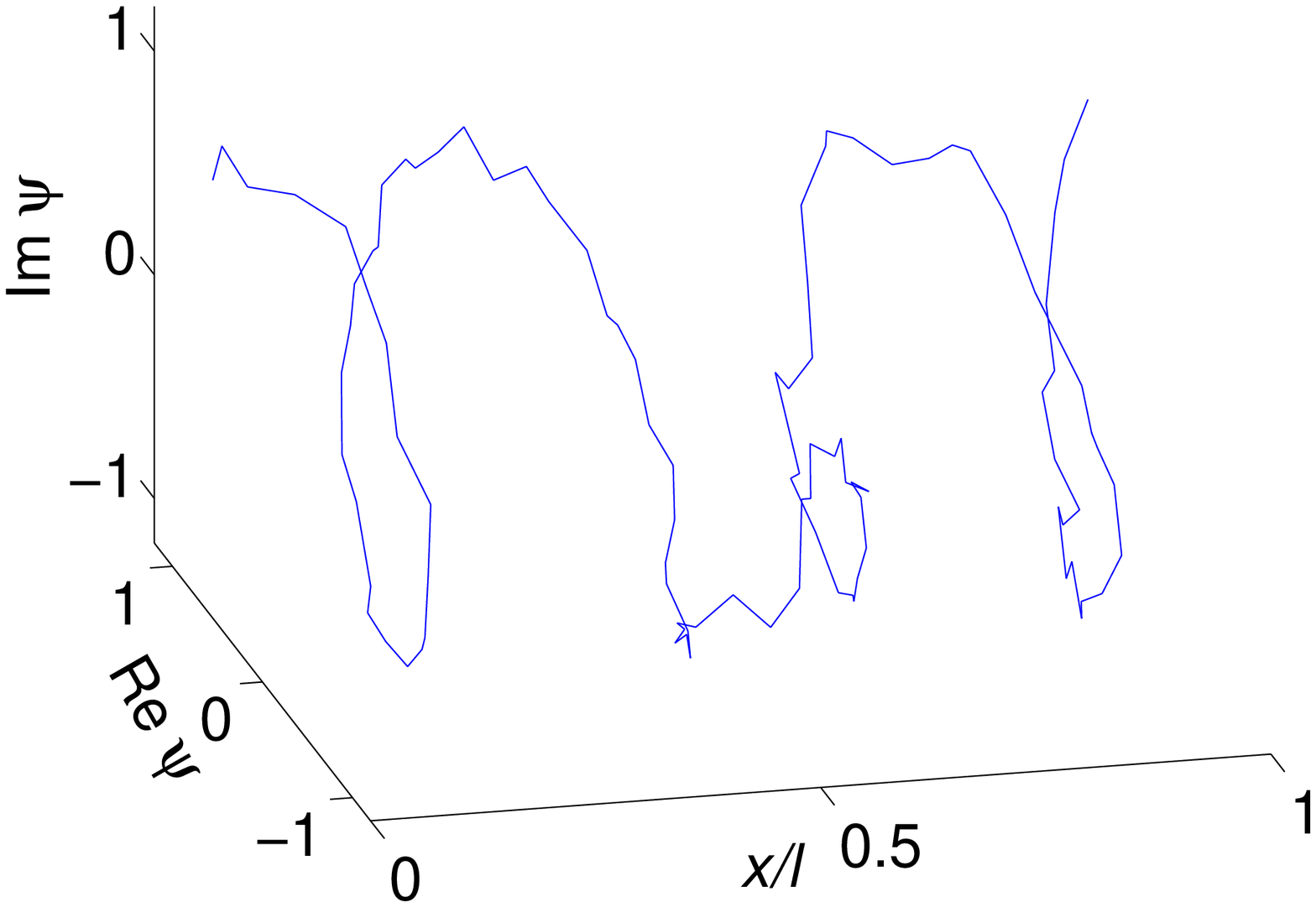,height=2.6cm} \hspace{1.5cm}
 \psfig{figure=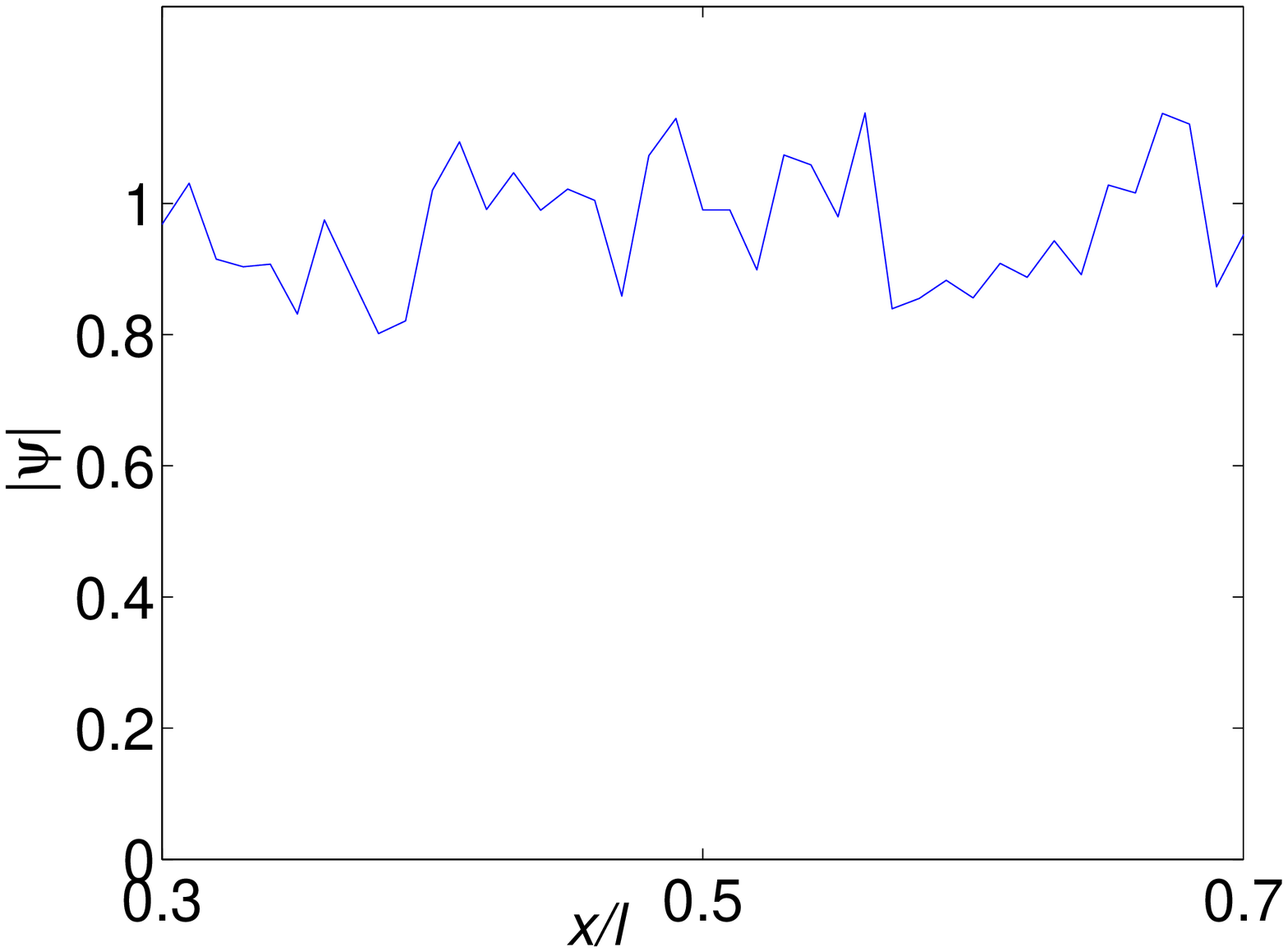,height=2.6cm}}
\centerline{\psfig{figure=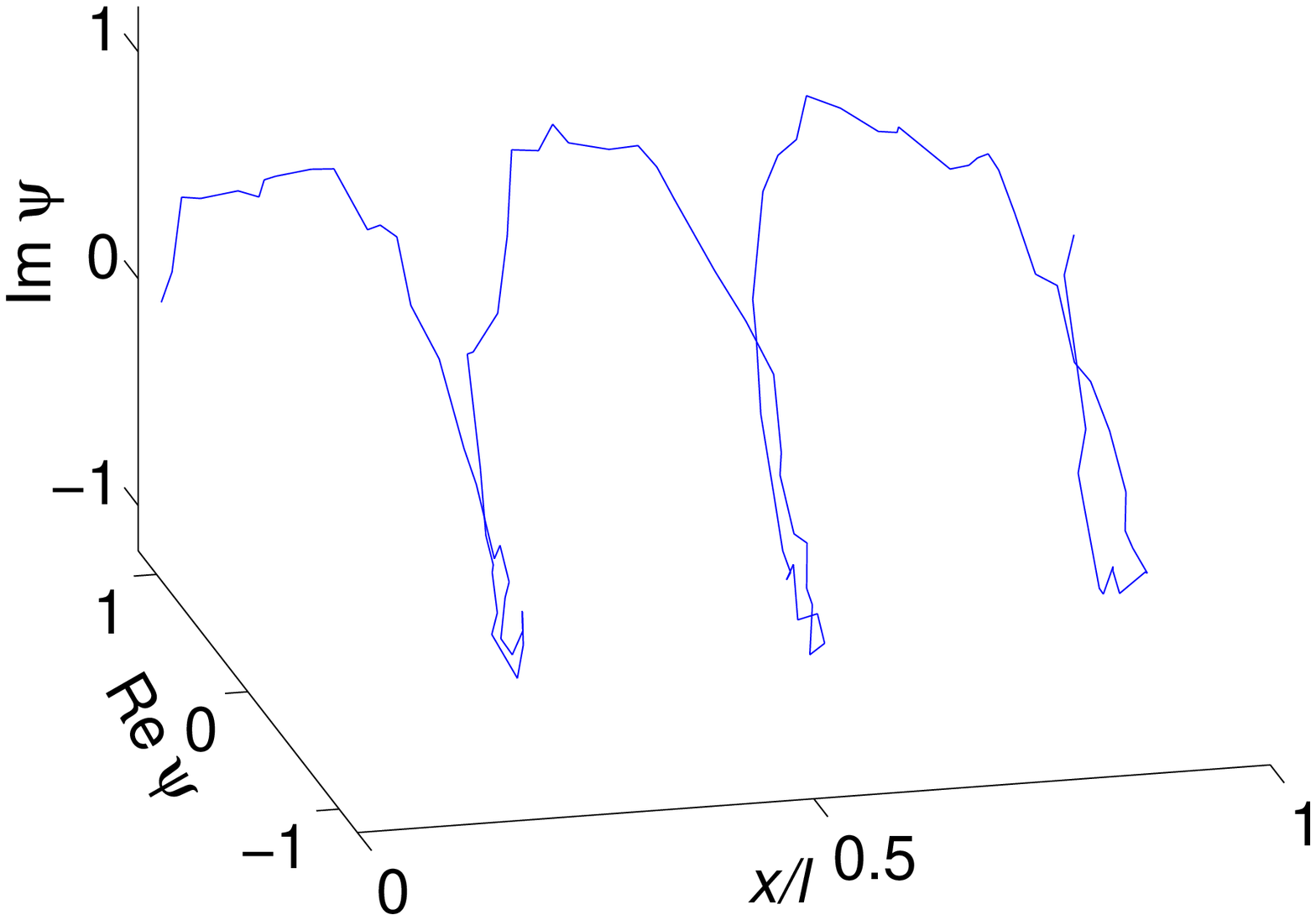,height=2.6cm} \hspace{1.5cm}
 \psfig{figure=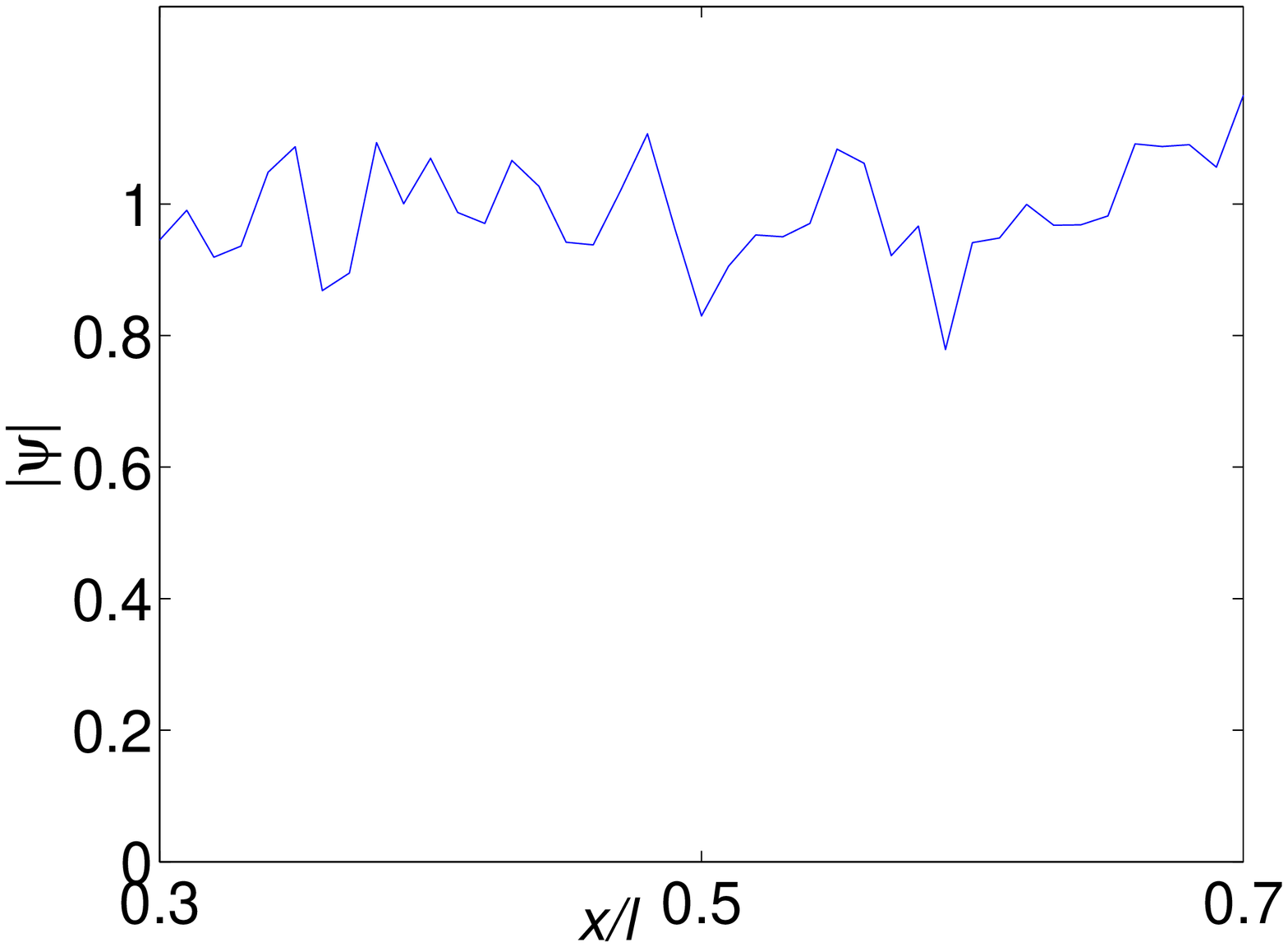,height=2.6cm}}
\bigskip
\caption{Transition pathway obtained from stochastic simulation,
from $\psi_4$ to $\psi_3$ with thermal noise. 
Left column: $(Re \psi,Im \psi)$ as a function of $x$; 
Right column: $|\psi|$ as a function of $x$. 
The figures at the top and bottom correspond to $\psi_4$ and $\psi_3$, 
respectively. The third from the top is closest to the saddle point.
In the fourth figure, $|\psi|$ reaches zero in the middle.}
\label{fig-path43}
\end{figure}

For comparison, we have carried out direct simulations for the motion 
of $\psi$ in the presence of thermal noise, using the stochastic equation
(\ref{TDGLE2}). For $k_BT\ll\sigma\xi\alpha_0^2(T_c-T)^2/\beta$, 
reasonably clean transition pathways can be obtained from the rare
transition events which carry the system from one metastable state to 
the other. Figure \ref{fig-path43} displays a sequence of $\psi(x)$ and 
$|\psi(x)|$, collected along a transition pathway from 
$\psi_4(x)$ to $\psi_3(x)$, calculated for $l=32\pi$ and 
$k_BT=0.02\sigma\xi\alpha_0^2(T_c-T)^2/\beta$.
A comparison of Figs. \ref{fig-MEP43} and \ref{fig-path43} shows
remarkable similarities. The advantage of a MEP is also seen from this
comparison: As a smooth path in configuration space, the MEP
reveals the transition behavior better than those noisy pathways obtained
from stochastic simulations. While some fine features of the transition
may be lost due to the noise in stochastic simulations, they can be well
preserved in the MEP. In particular, in order to obtain a clean
pathway from stochastic simulation, the temperature must be kept low
enough to reduce local fluctuations, but a low temperature inevitably
makes the transition events rare and difficult to catch, thus requiring
very long simulation time. 

\begin{figure}
\centerline{\psfig{figure=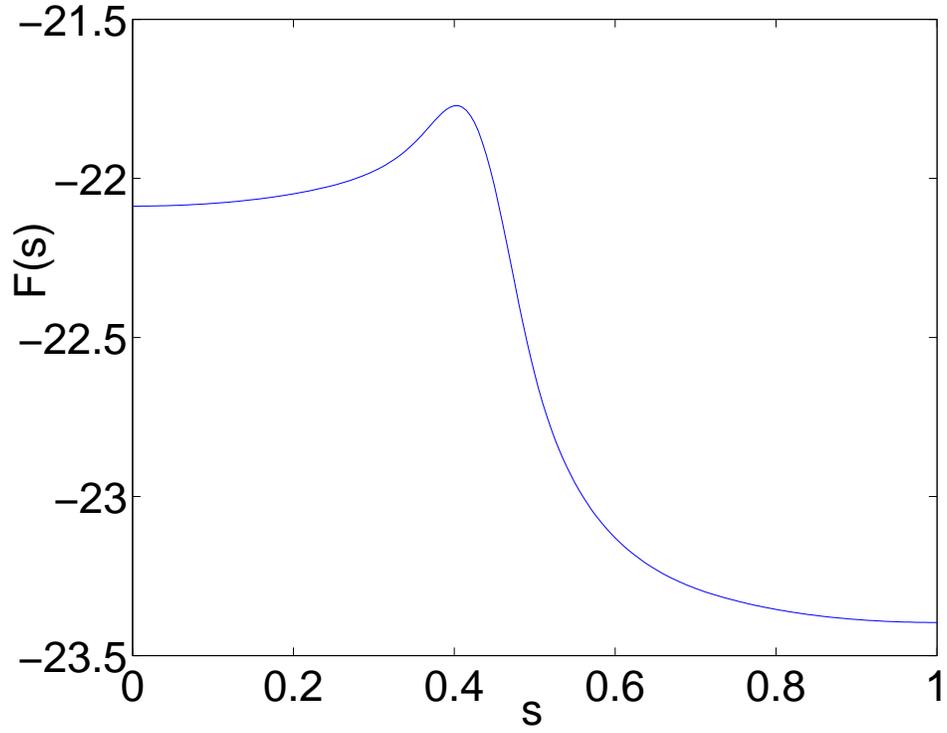,height=10cm}}
\bigskip
\caption{Scaled free energy $F$ evaluated along the MEP from 
$\psi_4$ to $\psi_3$, plotted as a function of the arc length $s$
in the $\psi(x)$-function space. The $\psi_4$ state is taken as 
the reference point where $s=0$.} \label{fig-ener43}
\end{figure}

\begin{figure}
\centerline{\psfig{figure=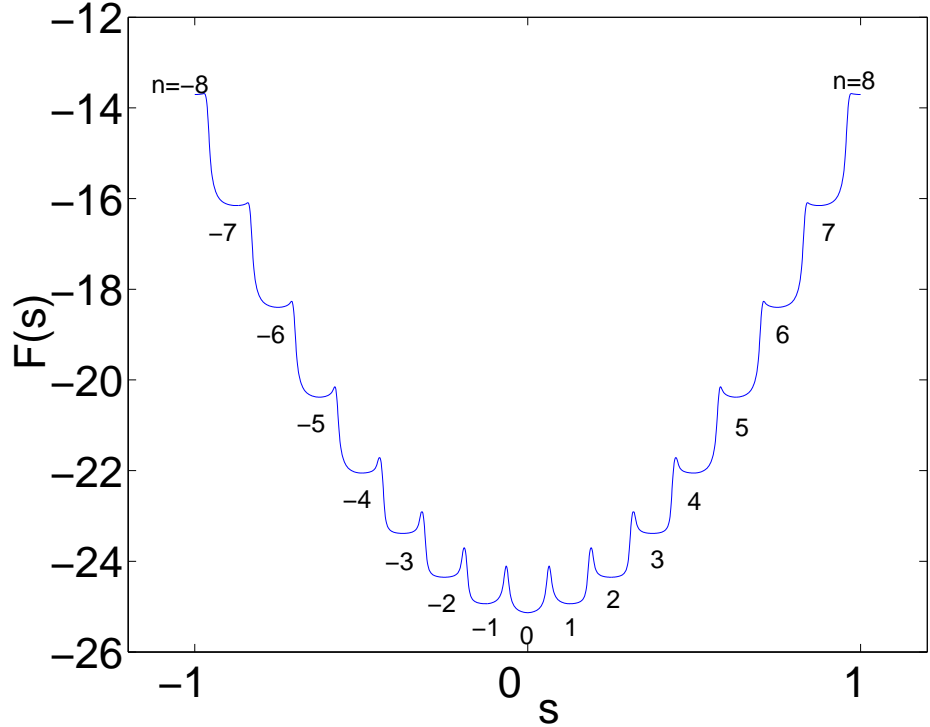,height=10cm}}
\bigskip
\caption{Scaled free energy $F$ evaluated along the MEPs 
connecting a sequence of metastable ($n\ne 0$) and 
stable ($n=0$) states, plotted as a function of the arc length $s$ 
in the $\psi(x)$-function space.
The scaled system length is $l=32\pi$ and the zero-current state 
($n=0$) is taken as the reference point where $F=-8\pi$ and $s=0$.
The maximum $|n|$ allowed by $|k_n|<k_c$ is $9$. The plot here
is up to $n=\pm 8$.}\label{fig-energy}
\end{figure}

Figure \ref{fig-ener43} shows the energy variation along the MEP 
from $\psi_4$ to $\psi_3$. The dimensionless free-energy barrier 
for the transition $\psi_4\rightarrow\psi_3$ is obtained as
$\Delta F_-=F[\psi_s]-F[\psi_4]=0.338$. 
Figure \ref{fig-energy} shows the energy profile along the MEP
from $\psi_{8}$ to $\psi_{-8}$. This MEP consists of $16$ segments, each
connecting two neighboring metastable states $\psi_n$ and $\psi_{n-1}$,
with $n$ running from $8$ to $-7$.

\subsection{Prefactor}\label{prefactor}

In calculating the prefactor $\Omega_-$ in Eq. (\ref{real-omega}) for $n=4$, 
we use the following procedure: 

(1) From the MEP calculated in Sec. \ref{mep}, the minimum 
${\mbox{\boldmath$\eta$}}_n$ and the saddle point 
${\mbox{\boldmath$\eta$}}_s$ are obtained.

(2) The (unphysical) degenerate directions ${\bf u}_n^{(1)}$ 
at ${\mbox{\boldmath$\eta$}}_n$ and ${\bf u}_s^{(2)}$ at
${\mbox{\boldmath$\eta$}}_s$ in the ${\mbox{\boldmath$\eta$}}$-function 
space are obtained by simple rotation according to 
Eqs. (\ref{un1}) and (\ref{us2}).

(3) The degenerate direction ${\bf u}_s^{(3)}$ at 
${\mbox{\boldmath$\eta$}}_s$ in the ${\mbox{\boldmath$\eta$}}$-function 
space is calculated using Eq. (\ref{us3}), and the parameter $\Lambda$ 
defined in Eq. (\ref{dcs3}) is then obtained to be $0.894$.

(4) The unstable direction ${\bf u}_s^{(1)}$ at $\mbox{\boldmath$\eta$}_s$ 
is obtained from the normalized difference between 
two neighboring configurations, evaluated at the saddle point along the MEP.
The corresponding negative eigenvalue is obtained as 
$\lambda_s^{(1)}=\langle{\bf u}_s^{(1)},H_s{\bf u}_s^{(1)}\rangle=-0.364$;

(5) The Hessians $H_n$ and $H_s$ are modified to give two 
positive definite matrices
\begin{equation} \label{eqn-modifyHn}
\tilde H_n= H_n + \nu_n^{(1)}\left[{\bf u}_n^{(1)}\right]
\left[{\bf u}_n^{(1)}\right]^T
\end{equation} 
and
\begin{equation} \label{eqn-modifyHs}
\tilde H_s= H_s + \sum _{i=1}^{3}
\nu_s^{(i)}\left[{\bf u}_s^{(i)}\right]\left[{\bf u}_s^{(i)}\right]^T
\end{equation}
where $\nu_n^{(1)}$ and $\nu_s^{(i)}$'s are all positive parameters;
In our calculation the parameters $\nu_n^{(1)}$ and $\nu_s^{(i)}$'s
are set to be $1$.

\begin{figure}
\centerline{\psfig{figure=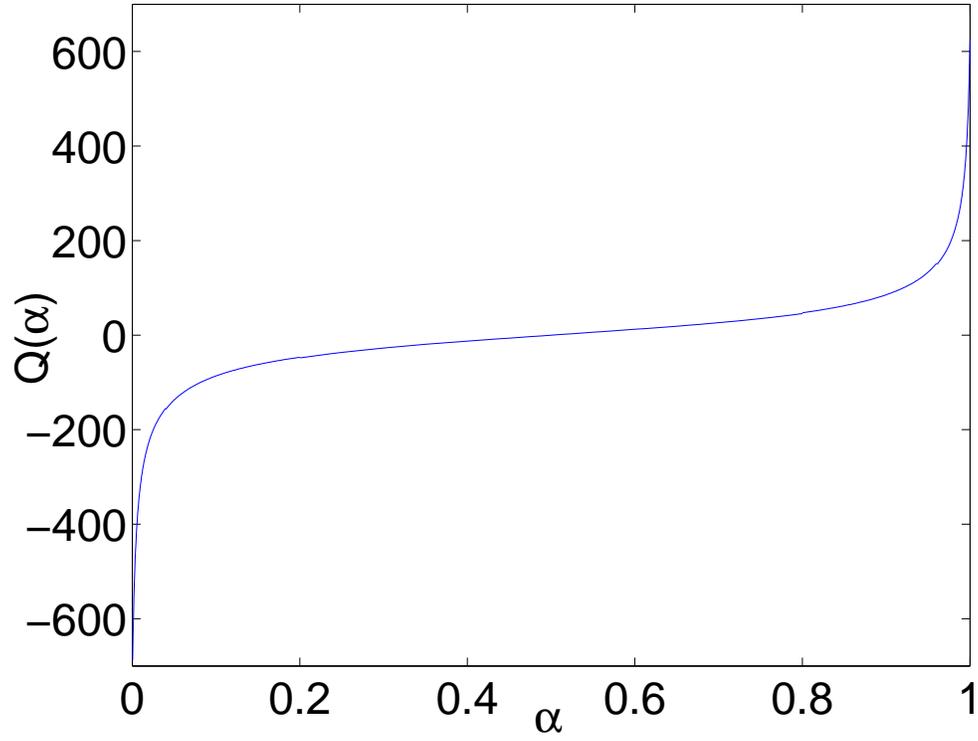,height=10cm}}
\bigskip
\caption{The expectation value $Q(\alpha)$ defined in Eq. (\ref{expecta}),
plotted as function of $\alpha$ in the interval $[0,1]$.}
\label{fig-Q-alpha}
\end{figure}

(6) The ratio $\det  \tilde H_n/ \det  \tilde H_s$ is calculated using 
Eq. (\ref{determinant-ratio}). Figure \ref{fig-Q-alpha} shows
the expectation value $Q(\alpha)$, defined in Eq. (\ref{expecta}) 
as a function of $\alpha$ in the interval $[0,1]$. This interval of
$\alpha$ is discretized using a non-uniform mesh of 352 points.
Since $Q(\alpha)$ varies rapidly near $\alpha=0$ and $1$, 
more points are distributed near these two ends, with the grid size 
$\Delta\alpha=5\times 10^{-4}$. In the middle of $[0,1]$, a larger 
grid size $\Delta\alpha =1.25\times 10^{-2}$ is used. For each $\alpha$, 
the stochastic equation (\ref{eqn-sde}) is simulated with 
$\epsilon=1$, and $Q(\alpha)$ is obtained from a time average
over $10^5$ realizations. 
The calculated value of the ratio $\det\tilde H_4/\det\tilde H_s$ 
is $0.608$. Using this ratio and the computed negative eigenvalue 
at the saddle point, we obtain the ratio $\det' H_4/|\det''H_s|=1.06$. 
Other sets of values have also been used for $\nu_n^{(1)}$ and
$\nu_s^{(i)}$ in Eqs. (\ref{eqn-modifyHn}) and (\ref{eqn-modifyHs}), 
and $\epsilon$ in Eq. (\ref{expecta}), but the final result of 
$\det' H_4/|\det''H_s|$ is not sensitive to those values.

\subsection{Rate}\label{rate}

Using $l=32\pi$ and $\sigma\xi\alpha_0^2(T_c-T)^2/\beta=
\sigma\xi H_c^2(T)/2\pi=50k_BT$ (used in stochastic simulation),
and the numerical values of $\Delta F_-=F[\psi_s]-F[\psi_4]=0.338$,
$\lambda_s^{(1)}=-0.364$, $\Lambda=0.894$, and 
$\det' H_4/|\det''H_s|=1.06$ obtained in Secs. \ref{mep} and 
\ref{prefactor}, we are ready to compute the rate 
(in the unit of $1/\tau(T)$)
\begin{equation}\label{rate-value}
\Gamma_-=\displaystyle\frac{|\lambda_s^{(1)}|}{4\pi^2}
\sqrt{\displaystyle\frac{\sigma\xi H_c^2(T)}{k_BT}}\Lambda l
\left[\displaystyle\frac{\det' H_4}{|\det'' H_s|}\right]^{1/2}
\exp\left[-\displaystyle\frac{\sigma\xi H_c^2(T)}{2\pi k_BT}
\Delta F_-\right],
\end{equation}
for the transition $\psi_4\rightarrow\psi_3$. 
The exponential factor is $e^{-0.338\times 50}\approx 4.6\times 10^{-8}$.
The prefactor $\Omega_-$ is evaluated as
$$\Omega_-=\displaystyle\frac{|-0.364|}{4\pi^2}\times\sqrt{2\pi\times 50}
\times 0.894\times 32\pi\times (1.06)^{1/2}\approx 15.1.$$
It follows that the dimensionless rate $\Gamma_-$ is approximately
$7\times 10^{-7}$. This value is in reasonable agreement with what
has been estimated through stochastic simulations.

\section{Discussion}

In this paper we have demonstrated that by using the string method,
thermal transition rates as formulated in the LAMH theory can be
numerically evaluated, even at low temperatures. In particular, the
pre-exponential factor may also be determined to some precision. Thus
the "electrical resistance" of a 1D superconductor may be evaluated
quantitatively. However, it has to be pointed out that quantum
tunneling effect, which can be important at low temperatures, is not
taken into account in the present formulation. Work is presently
underway to show that quantum tunneling can be similarly treated through
the string method, thus enabling a complete quantitative account of the
current dissipation phenomenon in 1D superconductors.

\section*{Acknowledgments}
This work was partially supported by Hong Kong RGC Grant HKUST6073/02P.

\appendix

\section{Langer-Ambegaokar Free-Energy Saddle Point}\label{app_LA}
By definition, the the lowest saddle point $\psi_s(x)$ between 
two neighboring metastable states $\psi_n$ and $\psi_{n-1}$ 
satisfies the stationary Ginzburg-Landau equation 
(\ref{stationaryGLE}). Langer and Ambegaokar have obtained
\begin{equation}\label{barrier-state}
\psi_s(x)=\left\{\sqrt{1-3k_s^2}\tanh\left[\sqrt{\displaystyle\frac
{1-3k_s^2}{2}}x\right]-i\sqrt{2}k_s\right\}e^{ik_sx},
\end{equation}
where $k_s$ is a wave vector determined by the condition 
$$
\phi(l/2)-\phi(-l/2)=k_sl+2\tan^{-1}\left(\displaystyle\frac
{\sqrt{1-3k_s^2}}{\sqrt{2}k_s}\right)=2\pi n,
$$
satisfying $k_{n-1}<k_s<k_n$.
Note that the amplitude of $\psi_s(x)$ is diminished 
in a small region around $x=0$, the phase-slip center.
Here we note that translating the phase-slip center from $x=0$
to $x_c$ will produce another saddle point $\psi_s(x-x_c)$,
because the free energy of the system is translationally invariant.
From the explicit expressions for $\psi_n$ and $\psi_s$, 
the dimensionless energy barriers can be readily obtained:
\begin{equation}\label{barrier-energy}
\begin{array}{cc}\Delta F_-=\displaystyle\frac{1}{4}
\left[\displaystyle\frac{8\sqrt{2}}{3}\sqrt{1-3k_s^2}
-8k_s(1-k_s^2)\tan^{-1}\displaystyle\frac
{\sqrt{1-3k_s^2}}{\sqrt{2}k_s}\right],\\ 
\Delta F_+=\displaystyle\frac{1}{4}
\left[\displaystyle\frac{8\sqrt{2}}{3}\sqrt{1-3k_s^2}
+8k_s(1-k_s^2)\left(\pi-\tan^{-1}\displaystyle\frac
{\sqrt{1-3k_s^2}}{\sqrt{2}k_s}\right)\right].
\end{array}
\end{equation}
Here $\Delta F_-=F[\psi_s]-F[\psi_n]$ and 
$\Delta F_+=F[\psi_s]-F[\psi_{n-1}]$.
Since $\Delta F_-<\Delta F_+$ (for $k_s>0$), the transition 
from $\psi_n$ to $\psi_{n-1}$ is more probable than that 
from $\psi_{n-1}$ to $\psi_n$. As a consequence,
thermally activated phase slips are current-reducing
dissipative process.

\section{McCumber-Halperin Expression for $\Lambda$ }\label{app_MH}
By writing the real and imaginary parts as the two components of 
a vector, the saddle-point state $\psi_s$ in Eq. (\ref{barrier-state}) 
can be written as
\begin{equation}\label{barrier-state-vector}
\begin{array}{ll}
{\mbox{\boldmath$\eta$}}_s(x) & =\left[\begin{array}{cc}
\cos k_sx & -\sin k_sx \\ \sin k_sx & \cos k_sx \end{array}\right]
\left[\begin{array}{c} \sqrt{1-3k_s^2}\tanh\left[\sqrt{\displaystyle\frac
{1-3k_s^2}{2}}x\right] \\ -\sqrt{2}k_s \end{array} \right]\\ 
& =\left[\begin{array}{cc}
\cos k_sx & -\sin k_sx \\ \sin k_sx & \cos k_sx \end{array}\right]
\tilde{\mbox{\boldmath$\eta$}}_s(x),
\end{array}
\end{equation}
from which 
\begin{equation}\label{zero-space}
\begin{array}{ll}
\nabla{\mbox{\boldmath$\eta$}}_s(x)=
\left[\begin{array}{cc}0 & -1 \\ 1 & 0 \end{array}\right]
{\mbox{\boldmath$\eta$}}_s(x)+\left[\begin{array}{cc}
\cos k_sx & -\sin k_sx \\ \sin k_sx & \cos k_sx \end{array}\right]
\nabla\tilde{\mbox{\boldmath$\eta$}}_s(x)
\end{array}
\end{equation}
is obtained. This equation indicates that while 
$\nabla{\mbox{\boldmath$\eta$}}_s$ is an eigenvector of $H_s$
corresponding to the zero eigenvalue, i.e., 
$H_s\nabla{\mbox{\boldmath$\eta$}}_s={\bf 0}$ by definition, it can be
decomposed into two physically distinct components. The first component, 
\begin{equation}\label{1st-component}
\left[\begin{array}{cc}0 & -1 \\ 1 & 0 \end{array}\right]
{\mbox{\boldmath$\eta$}}_s(x)=[\alpha_s^{(2)}]^{-1}{\bf u}_s^{(2)}(x),
\end{equation}
is the eigenvector corresponding to the zero eigenvalue 
$\lambda_s^{(2)}$, arising from gauge invariance. The second component,
\begin{equation}\label{2nd-component}
\left[\begin{array}{cc}
\cos k_sx & -\sin k_sx \\ \sin k_sx & \cos k_sx \end{array}\right]
\nabla\tilde{\mbox{\boldmath$\eta$}}_s(x)=
\Lambda{\bf v}_s^{(3)}(x),
\end{equation}
is the eigenvector corresponding to the zero eigenvalue 
$\lambda_s^{(3)}$, arising from translational symmetry.
Here a constant $\Lambda$ is introduced for normalization.
It is easily seen that the global rotation of the phase angle
is achieved by changing ${\mbox{\boldmath$\eta$}}_s$ 
in the direction of ${\bf u}_s^{(2)}$ while the translation
of the phase-slip center is achieved by changing 
${\mbox{\boldmath$\eta$}}_s$ 
in the direction of ${\bf v}_s^{(3)}$.
The normalization constant $\Lambda$ is determined by $\Lambda^2=
\int dx\left[\nabla\tilde{\mbox{\boldmath$\eta$}}_s(x)\right]^2
=\displaystyle\frac{2\sqrt{2}}{3}(1-3k_s^2)^{3/2}$.
The change of ${\mbox{\boldmath$\eta$}}_s(x)$ due to a small 
change in the location of the phase-slip center, $x_c$, is 
\begin{equation}\label{center-translation}
\left[\begin{array}{cc}
\cos k_sx & -\sin k_sx \\ \sin k_sx & \cos k_sx \end{array}\right]
\nabla\tilde{\mbox{\boldmath$\eta$}}_s(x)dx_c=
{\bf v}_s^{(3)}(x)dc_s^{(3)}. 
\end{equation}
Substituting Eq. (\ref{2nd-component}) into 
Eq. (\ref{center-translation}) yields $dc_s^{(3)}=\Lambda dx_c$
and thus $\int dc_s^{(3)}=\Lambda \int dx_c=\Lambda l$.
Therefore the sample length dependence arises naturally from 
the translational degeneracy. 
 
We want to point out that in general, ${\bf v}_s^{(3)}(x)$ is not 
orthogonal to ${\bf u}_s^{(2)}(x)$. From the inner products
\begin{equation}\label{inner-product1}
\begin{array}{ll}
&\int dx \left[\left[\begin{array}{cc}0 & -1 \\ 1 & 0 \end{array}\right]
{\mbox{\boldmath$\eta$}}_s(x)\right]^T
\left[\begin{array}{cc}0 & -1 \\ 1 & 0 \end{array}\right]
{\mbox{\boldmath$\eta$}}_s(x) \\
=&\int dx \left[{\mbox{\boldmath$\eta$}}_s(x)\right]^T
{\mbox{\boldmath$\eta$}}_s(x)
=(1-k_s^2)l-2\sqrt{2}\sqrt{1-3k_s^2},
\end{array}
\end{equation}
\begin{equation}\label{inner-product2}
\begin{array}{ll}
&\int dx \left[\left[\begin{array}{cc}
\cos k_sx & -\sin k_sx \\ \sin k_sx & \cos k_sx \end{array}\right]
\nabla\tilde{\mbox{\boldmath$\eta$}}_s(x)\right]^T
\left[\begin{array}{cc}
\cos k_sx & -\sin k_sx \\ \sin k_sx & \cos k_sx \end{array}\right]
\nabla\tilde{\mbox{\boldmath$\eta$}}_s(x) \\
=&\int dx \left[\nabla\tilde{\mbox{\boldmath$\eta$}}_s(x)\right]^T
\nabla\tilde{\mbox{\boldmath$\eta$}}_s(x)
=\displaystyle\frac{2\sqrt{2}}{3}(1-3k_s^2)^{3/2},
\end{array}
\end{equation}
\begin{equation}\label{inner-product3}
\begin{array}{ll}
&\int dx \left[\left[\begin{array}{cc}0 & -1 \\ 1 & 0 \end{array}\right]
{\mbox{\boldmath$\eta$}}_s(x)\right]^T
\left[\begin{array}{cc}
\cos k_sx & -\sin k_sx \\ \sin k_sx & \cos k_sx \end{array}\right]
\nabla\tilde{\mbox{\boldmath$\eta$}}_s(x) \\
=&2\sqrt{2}k_s\sqrt{1-3k_s^2},
\end{array}
\end{equation}
we obtain
\begin{equation}\label{inner-product4}
\int dx \left[{\bf u}_s^{(2)}(x)\right]^T{\bf v}_s^{(3)}(x)=
\displaystyle\frac{3^{1/2}8^{1/4}k_s}
{\left[(1-k_s^2)l-2\sqrt{2}\sqrt{1-3k_s^2}\right]^{1/2}
(1-3k_s^2)^{1/4}},
\end{equation}
which approaches zero as $k_s\rightarrow 0$ and/or
$l\rightarrow \infty$. So the two are orthogonal only in these limits.

\newpage

\newpage

\newpage

\newpage
\end{document}